\documentclass{sig-alternate-05-2015}
\CopyrightYear{2016} \setcopyright{acmcopyright}
\conferenceinfo{To appear at SIGMETRICS '16,}{June 14--18, 2016, Antibes Juan-Les-Pins, France.}
\isbn{978-1-4503-4266-7/16/06}\acmPrice{\$15.00}
\doi{http://dx.doi.org/XXXX.XXXX}

\usepackage{amsfonts}
\usepackage{eucal,color,comment}
\usepackage{amssymb,epsfig,latexsym}
\usepackage{amsmath,amssymb,latexsym,psfrag}
\usepackage{graphics,graphicx}
\usepackage{xspace,epsfig,url}
\newcommand{\cA}{\mathcal{A}}
\newcommand{\cB}{\mathcal{B}}

\newcommand{\cZ}{\mathcal{Z}}
\newcommand{\cV}{\mathcal{V}}
\newcommand{\R}{\mathbb{R}}

\newcommand{\EE}{\mathbb{E}}
\newcommand{\PP}{\mathbb{P}}

\newcommand{\N}{\mathbb{N}}

\newtheorem{Theorem}{Theorem}[section]

\newtheorem{Lemma}[Theorem]{Lemma}

\def\ind{{\rm 1\hspace{-0.90ex}1}}

\newcommand{\diff}{{\rm\,d}}
\newcommand{\equaref}[1]{(\ref{eq:#1})}

\newcommand{\tgifeps}[3]{
\begin{figure}[htb]
\centering
\includegraphics[width=#1cm]{#2.eps}
\vspace{-3mm}
\caption{#3\label{fig:#2}}
\vspace{-2mm}
\end{figure}
}

\begin{document}
%

\title{Generalized threshold-based epidemics in random graphs: the power of extreme values 
}
%
%
%
%
%

\numberofauthors{3} 
%
\author{
%
%
\alignauthor
Michele Garetto\\
       \affaddr{University of Torino}\\
       \email{michele.garetto@unito.it}
\alignauthor
Emilio Leonardi\\
       \affaddr{Politecnico di Torino}\\
       \email{leonardi@polito.it}
\alignauthor Giovanni-Luca Torrisi\\
       \affaddr{IAC-CNR}\\
       \email{ torrisi@iac.rm.cnr.it}
}


\maketitle

\begin{abstract}
Bootstrap percolation is a well-known activation process in a graph, 
in which a node becomes active when it has at least $r$ active neighbors.
Such process, originally studied on regular structures, has been recently
investigated also in the context of random graphs, where it can serve as a simple
model for a wide variety of cascades, such as the 
spreading of ideas, trends, viral contents, etc. over large social networks.
In particular, it has been shown that in $G(n,p)$ the final active set 
can exhibit a phase transition for a sub-linear number of seeds. 
In this paper, we propose a unique framework to study similar
sub-linear phase transitions for a much broader class of graph models
and epidemic processes. Specifically, we consider i) a generalized version 
of bootstrap percolation in $G(n,p)$ with random activation thresholds 
and random node-to-node influences; ii) different random graph models,
including graphs with given degree sequence and graphs with
community structure (block model). The common thread of our work is to 
show the surprising sensitivity of the critical seed set size
to extreme values of distributions, which makes some systems dramatically 
vulnerable to large-scale outbreaks. We validate our results running simulation on 
both synthetic and real graphs.

\end{abstract}



\section{\!\!\!\!\! Introduction and related work}\label{sec:int}
Many fundamental phenomena occurring in various kinds of complex systems,
ranging from technological networks (e.g., transportation, communication, energy),  
to biological networks (e.g., neural, ecological, biochemical) and social networks
(in the real world or over the Internet) can be described by dynamical 
processes taking place over the underlying graph representing the system structure.
Such processes modify over time the internal state of nodes and spread 
across the network following the edges of the graph.

One of the most widely studied example of such dynamical processes is the 
epidemic process, which starts from an initial set of infected nodes (usually
referred to as seeds, chosen either deterministically or random) that can pass the infection
to other (susceptible) nodes (under many possible models), 
possibly causing a major outbreak throughout the network. 

In our work we consider a generalized model for the spreading of an 
\lq epidemic', in which nodes are characterized by an infection threshold $r$ (either 
deterministic or random), and become infected when they collect from their neighbors 
an amount of influence larger than $r$. A special case of our model
is the well known bootstrap percolation process, in which $r$ is an integer 
($r \geq 2$) and each edge exerts an influence equal to one: 
simply put, a node becomes infected when it has at least $r$ infected neighbors.

Bootstrap percolation has a rich history, having been initially proposed
in the area of statistical physics \cite{chalupa}. Due to its many physical 
applications (see \cite{adler} for a survey) it has been primarily studied over the years 
in the case of regular structures (lattices, grids, trees), most notably in a series of papers
by Balogh and Bollob\'{a}s (e.g., \cite{bollobas}).
More recently, bootstrap percolation has been investigated also in the context of random graphs,
which is the focus of this paper.
In our work we are especially interested in epidemics occurring 
on very large, irregular structures such as those representing friendship relationships
among people. This interest is motivated by the great popularity gained by 
online social platforms (e.g., Facebook, Twitter, Instagram, etc.), which, coupled
with the increasing availability of always-on connectivity through mobile personal
devices, has created an unprecedented opportunity for the rapid dissemination 
of various kinds of news, advertisements, viral videos, as well as a
privileged environment for online discussion, creation and consolidation of 
beliefs, political opinions, memes and many other forms of collective reasoning.     
In this respect, bootstrap percolation provides a simple, primitive
model that can be used to understand the diffusion of a generic \lq idea' which
requires a certain amount of \lq reinforcement' from neighbors to be locally adopted. 

Some results have already been obtained for particular random 
graph models. In particular, \cite{pittel} first considered bootstrap percolation 
in the random regular graph $G(n,d)$, while \cite{amini1} has
extended the analysis to random graphs with given vertex 
degrees (configuration model). The above two papers assume that 
node degree is either fixed \cite{pittel} or it has both finite expectation and finite second 
moment \cite{amini1}, implying that the cardinality of the 
seed set must scale linearly with $n$ to observe a non-negligible growth of the epidemics.
Both papers make use of the differential equation method to analyze the discrete Markov Chain 
associated with the epidemic process. The analysis in \cite{amini1} also allows
the threshold to vary among the nodes. 

A very different technique has been recently proposed in \cite{JLTV} to
study bootstrap percolation in Erd\"{o}s--R\'{e}nyi $G(n,p)$ graphs. 
This technique allows to analyze also scenarios in which
a sharp phase transition occurs with a number of seeds 
which is sublinear in $n$: below a critical seed set size, for which one
can get a closed-form asymptotic expression, the infection
essentially does not evolve, whereas above the critical size
$n - o(n)$ nodes get infected with high probability\footnote{Throughout this 
paper we shall use the following (standard) asymptotic notation.
Let $f,g:\R\to\R$ be two functions. We write: $f(x)=o(g(x))$  \textcolor{black}{or $f(x)\ll g(x)$  
and $g(x)=\omega(f(x))$ or $g(x)\gg f(x)$} if $\lim_{x\to\infty}\frac{f(x)}{g(x)}=0$;
$f(x)=O(g(x))$ if there exist $K>0$, $x_0\in\R$: $|f(x)|\leq K|g(x)|$, for any $x\geq x_0$;
$f(x)\sim g(x)$ if $\lim_{x\to\infty}\frac{f(x)}{g(x)}=1$.
Unless otherwise specified, in this paper all limits are taken as $n\to\infty$.}.
In $G(n,p)$, this behavior is possible only when the average node degree
itself grows with $n$ (i.e., $p \gg 1/n$). The technique proposed 
in \cite{JLTV} has been applied by \cite{amini2} to 
power-law random graphs  generated by the Chung-Lu model (with power law 
exponent $2 < \beta < 3$), obtaining the interesting result that, 
under bounded average node degree, a sublinear seed set size is enough to reach
a linear fraction of the nodes.

Also our work started from the approach proposed in \cite{JLTV},
which provides a simple and elegant way to explore phase transitions  
taking place at sub-linear scale.    
To operate at this scale, we let, if needed, the average node degree
to grow with $n$, since this can be considered an acceptable assumption
in many cases. Indeed, real social networks (and in particular 
online social networks), which evolve over time with the addition/removal 
of nodes/edges, often exhibit the so called densification 
phenomenon \cite{lesko}, meaning that the number
of edges grows faster than the number of nodes (hence the 
average degree grows with time)\footnote{in practice, asymptotic results 
provide very good predictions of what happens in large (but finite) systems whenever
the average degree is not too small, say significantly larger than $r$.}.


The main thread of our work is to show the high \lq vulnerability' (in terms 
of critical number of seeds) that arises in networks when we add 
inhomogeneities in any one of many possible ways (i.e., by adding variability in 
thresholds, edge weights, node degree, or network structure). Although
this effect has already been observed in epidemic processes, 
the way in which inhomogeneities affect bootstrap percolation 
can be so dramatic that just extreme values
of distributions (and not their particular shape) can determine the
critical size of the seed set. We believe that this result, which apparently has not been
recognized before, is of fundamental importance to better understand
the dynamics of epidemics in complex systems.  



\section{Notation and preliminaries}\label{sec:prel}
We start introducing some background material and notation taken from \cite{JLTV},
which is necessary to follow the rest of the paper. As already mentioned, \cite{JLTV}
provides a full picture of standard bootstrap percolation in 
Erd\"{o}s--R\'{e}nyi graphs $G(n,p)$. Nodes are characterized by a common integer 
threshold $r \geq 2$, and the process starts with an initial set $\cA(0)$  
of vertices (the seeds), of cardinality $a$, which are chosen uniformly at random
among the nodes. We will use the same terminology adopted in \cite{JLTV}, where
infected nodes are called \lq active', whereas non-infected nodes are said to be 
inactive. An inactive node becomes active as soon as at least $r$ of its  
neighbors are active.  Note that seeds are declared to be active irrespective of the 
state of their neighbors. Active nodes never revert to be inactive, so the set of active nodes
grows monotonically.
  
The bootstrap percolation process naturally evolves through generations 
of vertices that become active.
The first generation is composed of all those vertices which are activated by the seeds.
The second generation of active nodes is composed by all the nodes which are activated 
by the joint effect of seeds and first generation nodes, etc. The process stops when either 
an empty generation is obtained or all nodes are active.

Now, it turns out that there is a useful reformulation of the problem
that makes the process especially simple to analyze. This reformulation, which 
was originally proposed in \cite{scalia}, consists in changing the time scale, 
by introducing a virtual (discrete) time step $t \in \N$, such that 
a single active node is \lq explored' at each time step (if the process has not yet stopped).  
By so doing, we forget about the generations, obtaining a more amenable
process which is equivalent to the original one, in terms of the final size  
of the epidemic.%

The above reformulation requires to introduce, besides the set $\cA(t)$
of nodes which are active at time $t$, another set $\cZ(t) \subseteq \cA(t)$, 
referred to as {\em used} vertices, which is the subset of active
vertices, of cardinality $t$, explored up to time $t$.
More precisely, at time zero the set $\cA(0)$ is initialized to the seed set, 
while the set of used vertices is initialized to the empty set: $\cZ(0) =\emptyset$.
Each node $i$ is given a counter $M_{i}(t) \in \N$, initialized to 0 at time $t=0$.

At time $t=1$ we arbitrarily choose a node $z(1)\in \cA(0)$ and we \lq fire' its 
edges, incrementing by one the counter of all its neighbors. By so doing, we {\em use}
node $z(1)$, adding it to the set of used nodes, so that $\cZ(1)=\{z(1)\}$.
We continue recursively: at each time $t$, we arbitrarily select an active node 
which has not been already used, i.e., $z(t) \in \cA(t-1)\setminus \cZ(t-1)$, and we 
distribute new \lq marks' to its neighbors, which are not in $\cZ(t-1)$, incrementing their counters.  
Node $z(t)$ is added to the set of used vertices: $\cZ(t)=\cZ(t-1)\cup\{z(t)\}$.
We then check whether there are some inactive vertices, denoted by set $\Delta \cA(t)$, 
that become active for effect of the marks distributed at time $t$ (i.e., vertices 
whose counter reaches $r$ at time $t$). Such newly activated vertices are added to the set of active vertices: $\cA(t)=\cA(t-1)+\Delta \cA(t)$ (note that no vertices can be activated
at time 1, being $r \geq 2$).

The process stops as soon as $\cZ(t)=\cA(t)$, i.e. when all active nodes have been used.
Let $T=\min \{t: \cZ(t)=\cA(t)\}$. By construction, the final size $A^*$ of the epidemic
is exactly equal to $T$: $A^*:=|\cA(T)|=|\cZ(T)|=T$.

The above reformulation of the problem is particularly useful
because the counter associated to each inactive node can be expressed as:
\begin{equation}\label{eq:counterbasic}
M_i(t) = \sum_{s=1}^t I_i(s)
\end{equation}
i.e., as the sum of $t$ independent Bernoulli random variables $I_i(s)$ of 
average $p$, each associated with the existence/non existence
of an edge in the underlying graph, between the node used at time $s$  
and node $i$. Indeed, it is perfectly sound to \lq reveal' the edges going out of a node 
just when the node itself is used (principle of deferred decision).
Moreover we can, for convenience, express the counters of all of the nodes 
at any time $t \geq 1$ just like \equaref{counterbasic},
without affecting the analysis of the final size of the epidemics.
Indeed, by so doing we introduce extra marks that are not assigned 
in the real process (where each edge is revealed at most one, in a single direction),
specifically, when a used node is \lq infected back' by a neighboring
used node. However, this \lq error' does not matter, since
 {\textcolor{black}{ it has no impact on the percolation process}.  
Note that counters $M_i(t)$ expressed in such a way are independent from node to node.

The dynamics of the epidemic process are determined by the behavior of the number $A(t)$ of \lq usable'
nodes (i.e., active nodes which have not been already used):
\[
A(t)=|\cA(t) \setminus \cZ(t)|= a -t + S(t)
\]
where $S(t)$ represents the number of vertices, which are not in the original seed set, that 
are active at time $t$. Note that the final size of the epidemics equals the first time $T$
at which $A(T) = 0$. Moreover, by construction, the number of used vertices at time 
$t$ equals $t$.
Now, let $\pi(t) := \PP(M_1(t) \geq r) = \PP(\textrm{Bin}(t,p) \geq r)$
be the probability that an arbitrary node not belonging to the seed set 
is active at time $t$. There are $n-a$ such nodes, each active independently of others, 
hence $S(t) \in \textrm{Bin}(n-a,\pi(t))$.

In essence, we need to characterize trajectories of process $A(t)$ which, besides
a deterministic component $a-t$ (decreasing with time), includes a random variable $S(t)$ which
is binomially distributed, with time-dependent parameter $\pi(t)$ (increasing with time):
\begin{equation}\label{eq:2bin}
A(t)= a -t + \textrm{Bin}(n-a,\PP(\textrm{Bin}(t,p) \geq r)) 
\end{equation}
In particular, whenever we can prove that, for a given $t$, 
$\PP((\inf_{\tau\le t} A(t)) < 0) \to 0$, then we can conclude that at least 
$t$ vertices get infected w.h.p. Similarly, if, for a given $t$, 
$\PP(A(t)<0)\to 1$, we can conclude that the percolation terminates 
w.h.p. before $t$, thus the final  
number of infected vertices will be smaller than $t$.
We now present a simplified form of the main theorem in \cite{JLTV},
together with a high-level description of its proof.
\begin{Theorem}[Janson \cite{JLTV}]\label{thm:simplifiedjanson}
Consider bootstrap percolation in $G(n,p)$ with $r \geq 2$, and a number $a$ of seeds
selected uniformly at random among the $n$ nodes. Let $p = p(n)$ be such that
$p = \omega(1/n)$, $p=o(n^{-1/r})$. 
Define:
\begin{eqnarray}
t_c & := & \left( \frac{(r-1)!}{n p^r} \right)^{1/(r-1)} \label{eq:tcjanson} \\
a_c & := & \left(1 - \frac{1}{r}\right) t_c \label{eq:acjanson} 
\end{eqnarray}
If $a/a_c\to\alpha < 1$ (subcritical case), then w.h.p. the final size is $A^* < 2a$.
If $a/a_c\to\alpha \geq 1+\delta$, for some $\delta > 0$ (supercritical case), then w.h.p. $A^* = n - o(n)$.
\end{Theorem}
Note that, under the above assumptions on $p(n)$, the \lq critical time' $t_c$ is
such that both $t_c = \omega(1)$ and $t_c = o(n)$, and the same holds for
the critical number of seeds $a_c$, which differs from $t_c$ just by the constant factor
$(1-1/r)$, i.e., we get a phase transition for a sublinear number of seeds. 
 
The methodology proposed in \cite{JLTV} to obtain the above result
is based on the following idea: 
$A(t)$ is sufficiently concentrated around its mean that we can approximate it as
$A(t) \approx \EE(A(t)) = a - t + (n-a) \pi(t)$.
\textcolor{black}{
Now, for a wide range of values of $t$ (i.e., whenever $pt \to 0$, and in 
particular around $t_c$), $\pi(t)$ can be expressed as $\pi(t) = \frac{t^r p^r}{r!}(1 + O(pt + t^{-1}))$.
Therefore function $\EE(A(t))$ 
has a clear trend: it starts from $a$ at $t=0$ and first decreases
up to a minimum value reached at $t \approx t_c$, after
which it grows to a value of the order of $n$.
Hence, time $t_c$ acts as a sort of 
bottleneck: if $\EE(A(t_c))$ is positive (negative), we are in the supercritical 
(subcritical) case.
Finally, we can compute the asymptotic value of $t_c$ by finding the minimum of 
function $f(t) = n \frac{t^r p^r}{r!} - t$.
}
 
The result then follows considering that, starting from $a$ seeds, 
we get $\EE(A(t_c)) = a - a_c + o(a_c)$, and that by changing $a$ we deterministically
move up or down the process $A(t)$. Hence, if we assume that $a/a_c$ is asymptotically 
bounded away from 1 we obtain a sufficient \lq guard factor' around the trajectory of the mean process to conclude that the real process is either supercritical or subcritical
(see Fig. \ref{fig:janson}).

\tgifeps{6}{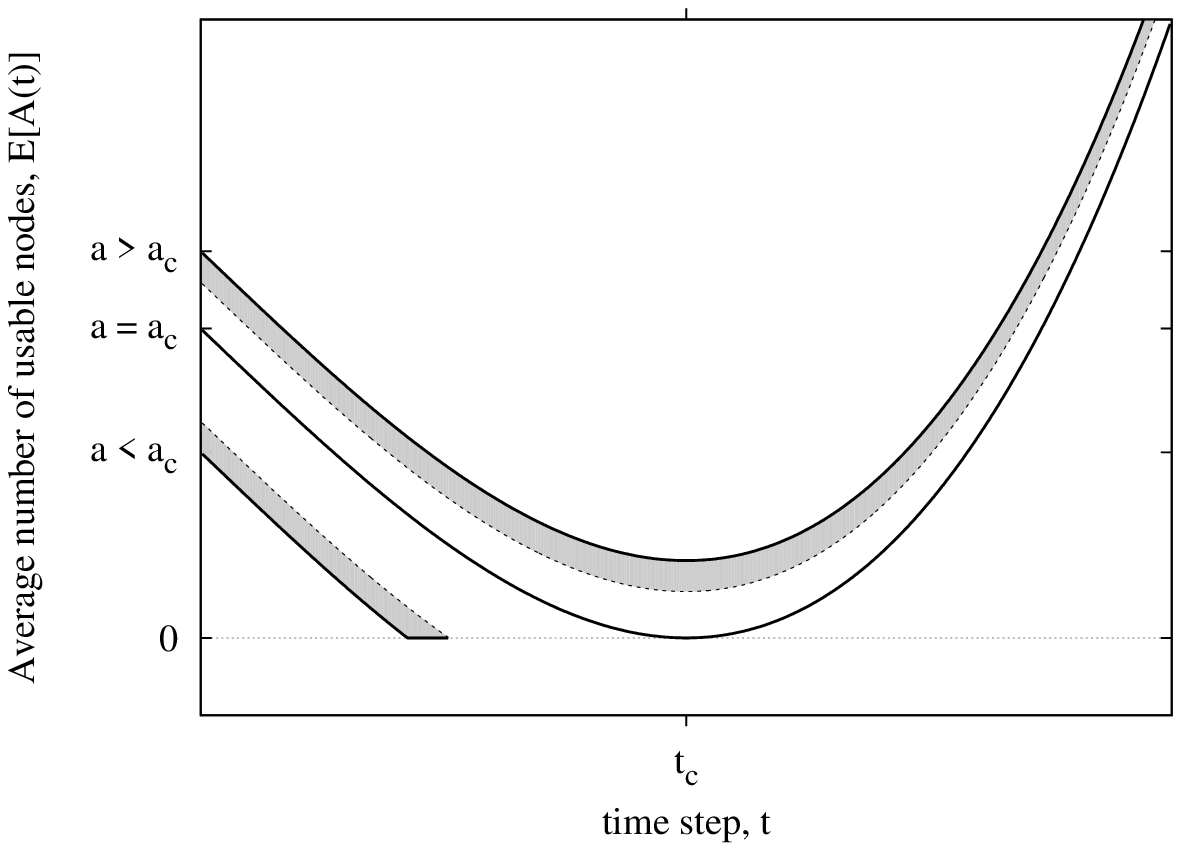}{Example of (asymptotic) trajectories of the mean number of usable nodes, 
$\EE(A(t))$, with $r=3$. The plot also illustrates by shaded regions the concept of \lq guard zone'.
}

We emphasize that in \cite{JLTV} authors use a 
martingale approach to show that $A(t)$ is sufficiently concentrated around its 
mean, which allows them to establish their results w.h.p.

As last premise, it is better to clarify why we assume $r \geq 2$.
The reason is that the case in which a node can be infected by just a single neighbor 
is degenerate, and leads to the trivial fact that a single seed is enough
to infect the entire connected components it belongs to.
Hence, one has to apply a totally different set of tools \cite{lof} to characterize the
final size of the epidemic. This case, however, is not interesting to us, since
the networks of many real systems are connected by construction, or they at least have
a giant connected component. Hence, no phase transitions occur here in 
the number of seeds.

\section{Summary of contributions}\label{sec:summary}
In this work we extend the approach of \cite{JLTV} along three \lq orthogonal' directions
that allow us to study more general threshold-based 
epidemic processes in inhomogeneous scenarios. 

\tgifeps{5}{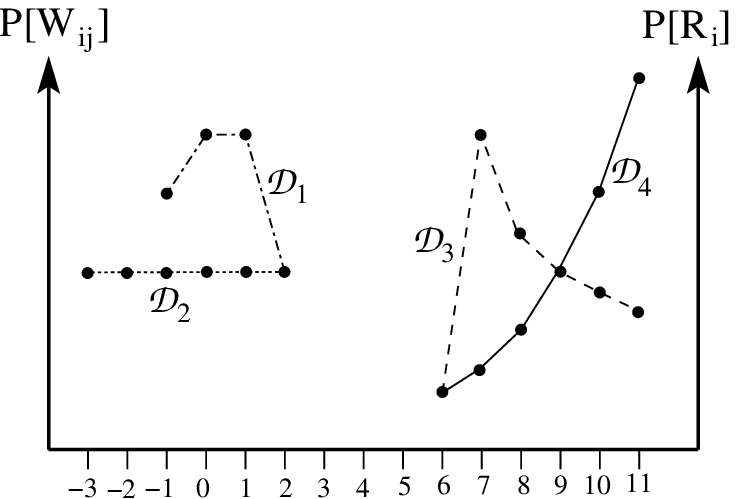}{Examples of distributions of $W_{ij}$ and $R_{i}$
leading to the same (asymptotic) critical number of seeds $a_c$.}

\begin{enumerate}
\item  We consider a generalized version of bootstrap percolation in $G(n,p)$, in which 
thresholds of nodes are i.i.d. random variables $R_i>0$, and infected
nodes transmit a random amount of infection to their neighbors. Specifically,
we assume that i.i.d. weights $W_{ij}$ are assigned to the edges of the graph,
representing the amount of infection transmitted through the edge. 
For this case, we obtain the asymptotic closed form expression of the 
critical number of seeds, and an exponential law for the probability that
the process is supercritical or subcritical, strengthening
the results in \cite{JLTV} (where results hold, instead, w.h.p.).
The most significant outcome of our analysis is that the critical number
of seeds typically {\color{black} does not depend on the entire distribution of $R_i$ and $W_{ij}$,
but just on values taken in proximity of the lower (for $R_i$) and
upper (for $W_{ij}$) extreme of their support.}
For instance, in Figure \ref{fig:esempio} we
show examples of two (discrete) distributions for $W_{ij}$, labelled ${\cal D}_1$
and ${\cal D}_2$, and two (discrete) distributions for $R_{i}$, labelled ${\cal D}_3$
and labelled ${\cal D}_4$. It turns out that any combination of them $({\cal D}_a,
{\cal D}_b)$, with $a \in \{1,2\}$ and $b \in \{3,4\}$ leads to the same
asymptotic critical number of seeds $a_c$. Note that the various
distributions have different means, and that one of them (${\cal D}_2$)
has even negative mean.

 
\item We extend the problem reformulation originally proposed in \cite{scalia}, where
a single node is used at each time, to a similar reformulation in which a single
edge is used at a time. This view is more convenient to apply 
the approach of \cite{JLTV} to other random graph models. In particular, we consider 
graphs with given degree sequence (configuration model), obtaining a closed-form
expression of the asymptotic critical number of seeds. We then compute
the scaling order of $a_c$ for the particular (but most significant) case of power-law 
degree sequence, considering a wider range of parameters with respect to the one 
studied by \cite{amini2}. Again, we observe the interesting phenomenon that in some cases
the precise shape of the degree distribution (i.e., the power law exponent) 
does not matter, since $a_c$ is determined by the largest degree.
 
\item  We extend the analysis to the so-called block model, which provides
a simple way to incorporate a community structure into a random graph model
while preserving the analytical tractability of $G(n,p)$.
We observe once more the interesting effect that the critical number of seeds
might be determined by a single entry of the matrix of inter- (or intra-) community 
edge probabilities (i.e., the most vulnerable community).
\end{enumerate}

Although we consider (for simplicity) the above three forms of 
inhomogeneity \lq in isolation', it is not particularly difficult to combine
them, if desired. Indeed, we show that all extensions above can be studied 
within a unique framework. 
We emphasize that in this paper we generally assume that seeds are selected
uniformly at random among the nodes, without knowledge of thresholds,
weights, degrees, network structure. This differentiates our analysis
from existing works addressing the so called influence
maximization problem, i.e., finding the seed set 
that maximizes the final size of the epidemic (e.g., \cite{kempe}).
{\color{black} We observe that in the influence
maximization framework many authors have already considered 
generalized models taking into account the impact
of edge weights, node-specific thresholds, etc. (e.g., variants of the linear
threshold model proposed in \cite{watts}). However, to the best of our
knowledge, \textcolor{black}{asymptotic properties of such generalized 
models are still not well understood.
This paper makes a step forward in this direction 
analysing sublinear phase-transitions occurring when seeds are allocated
uniformly at random in the network.
}}

Interestingly, in all cases that we consider 
the epidemic is triggered among the most vulnerable nodes, and then it spreads out 
hitting less and less vulnerable components of the network.
This fact can have dramatic consequences on the minimum number of seeds
that can produce a network-wide outbreak.  

In the following sections we present the above three 
contributions one at a time. Simulation experiments are presented along 
the way, to validate and better illustrate our analytical results.

   
\section{Generalized bootstrap percolation in $G(\MakeLowercase{n,p})$}\label{sec:gengnp}

\subsection{System model}\label{subsec:model}
We start considering Erd\"{o}s--R\'{e}nyi random graphs $G(n,p)$,
extending basic bootstrap percolation to the case in which node thresholds
and/or node-to-node influences are i.i.d random variables.
We denote by $R_i > 0$ the (real-valued) threshold associated to node $i$. 
We then assign a (real-valued) random weight $W_{ij}$ to each edge of the graph,
representing the influence that one node exerts on the other (see later).
Node $i$ becomes active when the sum of the weights on the edges
connecting $i$ to already active neighbors becomes greater than or equal to $R_i$.

Recall that each edge of the graph is \lq used' by the process at most once.
Hence our analysis encompasses both the \lq symmetric' case in which the influence 
(possibly) given by $i$ to $j$ equals the influence (possibly) given by $j$ to $i$,
and the \lq asymmetric' case in which weights along the two directions of an edge 
are different (i.i.d.) random variables. In both cases, we can consider a single
random weight on each edge.

We do not pose particular restrictions to the distributions of $R_i$
and $W_{ij}$, except for the following one, which avoids the degenerate case in which 
a node can get infected by a single edge (the case $r=1$ in basic bootstrap percolation):
\begin{equation} \label{eq:weightcond}
\text{ess} \inf R_i >  \text{ess} \sup W_{ij} > 0. 
\end{equation}

Note that we can also allow $W_{ij}$ to take negative values, which could represent,
in the context of social networks, neighbors whose behavior steers us away from the 
adoption of an idea. This generalization produces, indeed, rather surprising results, 
as we will see. However, negative weights require to introduce some extra
assumptions on the dynamics of the epidemics process, which are not needed
when weights are always non-negative. Specifically, with negative weights 
we must assume that i) once a node becomes infected, it remains infected for ever;
ii) some random delays are introduced in the infection process of 
a node and/or on the edges, to avoid
that a node receives the combined effect of multiple (simultaneous) influences
from active neighbors. We argue that assumption ii) is not particularly     
restrictive, since in many real systems influences received by a node take place
atomically (e.g., a user reading ads, posts, twits, and the like). 
Assumption i) instead is crucial, because with negative weights 
counters no longer increase monotonically, and thus they can traverse
the threshold many times in opposite directions. Assumption i) can be adopted,
however, to study many interesting epidemic processes whose dynamics are triggered
by nodes crossing the threshold for the first time\footnote{For example, 
on some online platforms, notifications that a user has watched a 
given viral video, bought a product, expressed interest for an event, etc.,
might be sent immediately (and once) to his friends, no matter if the
user changes his mind later on.}.

The analysis of the general case can be carried out by exploiting the same 
problem reformulation described in Sec. \ref{sec:prel}, in which a single
active node is used at each time step. Indeed, we can associate to inactive nodes
a (real-valued) counter, initialized to 0 at time $t=0$, which evolves according to: 
\begin{equation} \label{Mi}
 M_i(t)=\sum_{s=1}^{t}I_{i}(s)W_{i}(s)
\end{equation}
where $I_{i}(s)$, $\forall s,\forall i$, is a Bernoulli r.v. with average $p$ revealing the presence of edge $(z(s),i)$ and $W_{i}(s)$, $\forall s,\forall i$, 
is the random weight associated to the same edge.
Similarly to the basic case, the above expression of $M_i(t)$
can be extended to all nodes and all times, without affecting the results.
By so doing, counters $M_i(t)$ are independent from node to node.

We then re-define $\pi(t)$, as the probability that an arbitrary node
which is initially inactive (take node 1), has become active at any 
time $\tau \leq t$:
\[
\pi(t) := \PP(M_1(\tau)\geq R_1, \tau \leq t)
\] 
With the above definition, the system behavior is still determined
by trajectories of process \equaref{2bin}. We have:
\begin{align}
\pi(t)&= \PP\left(\sup_{\tau\le t} \sum_{s=1}^{\tau} I_1(s) W_1 (s) \geq R_1\right)\nonumber\\
&\stackrel{(a)}{=}\sum_{\rho=0}^{t}\binom{t}{\rho}p^\rho (1-p)^{t-\rho} \;\; \cdot \PP\left(\sup_{m \le \rho} \sum_{s=1}^m W_1(s) \geq R_1\right) \nonumber\\
&= \sum_{\rho=0}^{t} \binom{t}{\rho}p^\rho (1-p)^{t-\rho}q_\rho \label{eq:pibase}
\end{align}
where equation (a) is obtained by conditioning over the number $\rho$  
of variables $I_1(s)=1$, and we have defined $q_\rho$:
\[
q_\rho := \PP\left(\sup_{m\le \rho}\sum_{s=1}^{m} W_1(s) \geq R_1\right)
\]
which can be interpreted as the probability that a node, which has 
sequentially received the influence of $\rho$ infected neighbors, 
has become active. Let $q_{\infty} := \lim_{\rho \rightarrow \infty} q_\rho$.
Note that, as consequence of elementary properties of random walks,
$q_{\infty} = 1$ when $\EE[W_{ij}] \geq 0$ (recall also \equaref{weightcond}).
We introduce the following fundamental quantity:
\begin{equation}
\rho^* := \min\{\rho \ge 2:\,\,q_\rho>0\}
\end{equation}
In words, $\rho^*$ is the minimum number of infected neighbors
that can potentially (with probability $q_{\rho^*}$) activate a node. 
Note that, as consequence of \equaref{weightcond}, it must be $\rho^*\ge 2$.
For example, under the distributions shown in Fig. \ref{fig:esempio},
we have $\rho^* = 3$, $q_{\rho^*} = \PP(R_i = 6)\cdot\PP(W_{ij} = 2)^3$.

\subsection{Main results}\label{subsec:main}
We are now in the position to state our main results for
the generalized bootstrap percolation model in $G(n,p)$.
First, we define:
\[
t_c:=\left(\frac{(\rho^*-1!)}{np^{\rho^*}q_{\rho^*}}  \right)^{\frac{1}{\rho^*-1}};
\qquad a_c := \left( 1-\frac{1}{\rho^*}  \right)t_c 
\]
Moreover, we shall consider the function:
\begin{align}
H(x) &:=1-x+x\log x,\quad x>0,\quad H(0):=1, \nonumber \\
H(x)& :=+\infty,\quad x<0 \nonumber
\end{align}
\begin{Theorem}[Super-critical case]\label{thm:bpercolationsuper}
Under the assumptions: $1/(np)\to 0$, $p=o(n^{-1/\rho^*})$, $a/a_c\to\alpha$ for some $\alpha>1$.
Then, 
\[
\text{$\forall$ $\delta>0$},\quad \PP\left(\Big|\frac{A^*}{n}-q_\infty\Big|>\delta\right)=O\left(\mathrm{e}^{-C_1(\rho^*,\alpha)a+o(a)}\right),
\]
where $C_1(\rho^*,\alpha)$ is the constant:
\[
C_1(\rho^*,\alpha):= \min_{x\in[\alpha\frac{\rho^*-1}{\rho^*},\infty)} \frac{x^{\rho^*}}{\alpha(\rho^*-1)}
 H\left(\frac{x \rho^*  -\alpha(\rho^*-1)}{x^{\rho^*}}  \right).   
\]
\end{Theorem}

For the sub-critical case, we define the function \\
$h(x) := x-(\rho^*)^{-1}x^{\rho^*}-\alpha(1-(\rho^*)^{-1})$, 
for $x\in [0,1]$, $\alpha\in (0,1)$, and we
we denote by $\varphi(\alpha)$ the only\footnote{Function $h$ is continuous and 
strictly increasing on $[0,1]$ with \mbox{$h(0)=-\alpha(1-(\rho^*)^{-1})<0$}
and \mbox{$h(1)=(1-(\rho^*)^{-1})(1-\alpha)>0$}.} solution of $h(x)=0$, $x\in [0,1]$.
Furthermore, having defined the interval \mbox{$I:=\left(0,(1-\alpha)(1-(\rho^*)^{-1})\right)$}, 
it holds:
\begin{eqnarray*}
\hspace{-2mm} \forall \delta>0, \exists \varepsilon_\delta \in I: 
[-\delta,\delta]\supseteq [h^{-1}(-\varepsilon_\delta)-\varphi(\alpha),h^{-1}(\varepsilon_\delta)-\varphi(\alpha)]
\end{eqnarray*}
\begin{Theorem}[Sub-critical case]\label{thm:bpercolation}
Under the assumptions: $1/(np)\to 0$, $p=o(n^{-1/\rho^*})$ and $a/a_c\to\alpha$ for 
some $\alpha\in (0,1)$. Then, $\forall \delta>0$,
\begin{equation*}
\PP\left(\Big|\frac{A^*}{a}-\frac{\rho^*}{\rho^*-1}\frac{\varphi(\alpha)}{\alpha}\Big|>\delta\right)=O\left(\mathrm{e}^{-C_2\left(\rho^*,\alpha,\varepsilon_{\delta}\right)a+o(a)}\right),
\end{equation*}
where $\varepsilon_{\delta}$ and $\varphi(\alpha)$ are defined as above,
and
\[
C_2(\rho^*,\alpha,\varepsilon):=\frac{1}{\alpha(\rho^*-1)}H\left(1+\varepsilon\rho^*\right).
\]
\end{Theorem}

We shall provide here a sketch of the proof of Theorems 
\ref{thm:bpercolationsuper} and \ref{thm:bpercolation}.
The complete proofs, including all mathematical details, 
can be found in \cite{techrep}.

At high level, we can show that almost complete percolation occurs 
under super-critical conditions, by:
\begin{itemize}
\item[i)] analysing the trajectory of the mean of process \equaref{2bin},
$\mathbb E[A(t)]= a - t + (n-a) \pi(t)$, finding conditions under which the 
above quantity is positive (with a sufficient guard factor)  
for any $t< (q_\infty-\delta)n$, for arbitrarily small $\delta>0$.
\item[ii)]  showing that the actual process $A(t)$ 
is sufficiently concentrated around its mean that we can conclude that 
$A(t)>0$ w.h.p. for any  $t<(q_\infty-\delta)n$. 
\end{itemize}
For the sub-critical regime we can use similar arguments, 
showing that  $\mathbb E[A(t)]$ becomes negative at early stages, and that 
$A(t)$ is sufficiently concentrated around its average that we can claim 
that the actual process stops at early stages w.h.p.

We start from the asymptotic approximation of $\pi(t)$: 
\begin{equation} \label{eq:pi-summary}
\pi(t)=\frac{(pt)^{\rho^*}}{\rho^{*}!}(q_{\rho^*}+O(pt+t^{-1})).
\end{equation}
which holds for any $t$ such that $pt \to 0$.
The above approximation allows us to write, for any $t\ll p^{-1}$:
\[
\EE[A(t)]=a-t+(n-a)\pi(t) = a - t + n \frac{(pt)^{\rho^*}}{\rho^{*}!}q_{\rho^*}(1+o(1))
\]
under the further assumption that $a = o(n)$.
Thus, having defined for any $t\in \R_+$ 
function \mbox{$f(t)= a - t + \frac{(pt)^{\rho^*}}{\rho^{*}!} q_{\rho^*}$}, 
for $n$ large enough we can determinate the 
sign of $\EE[A(t)]$ for any $t\ll p^{-1}$ by analysing 
the behavior of $f(t)$.
Elementary calculus reveals that $f(t)$ has a unique minimum at: 
\[
t_c=\left(\frac{(\rho^*-1!)}{np^{\rho^*}q_{\rho^*}}  \right)^{\frac{1}{\rho^*-1}}
\]
with $f(t_c)=a-a_c$, $a_c = \left(1-\frac{1}{\rho^*}  \right)t_c$.
Thus, we obtain an asymptotic closed-form expression for the critical number of seeds $a_c$
(one can easily verify that, under the assumption $\frac{1}{n}\ll p \ll n^\frac{1}{\rho^*}$,
it holds $t_c \to \infty$, $a_c\to \infty$,  $pt_c \to 0$, $\frac{a_c}{n}\to 0$).



The difficult part of the proofs is to show that 
$A(t)$ is sufficiently concentrated around its expectation
that we can establish exponential bounds (as $n \to \infty$) 
on the probability that the final size of the epidemics falls outside the 
intervals stated in Theorems \ref{thm:bpercolationsuper} (super-critical case) 
and \ref{thm:bpercolation} (sub-critical case).

For the super-critical case, we adapt a methodology proposed in \cite{JLTV},
which separately considers four time segments\footnote{The boundaries of all segments are to be meant 
as integers. However, to simplify the notation, we will omit 
$\lfloor \cdot \rfloor$ and $\lceil \cdot \rceil$ symbols.}: 
i) segment\footnote{note that the process cannot stop at $t < a$.}
$[a,K t_c]$ (where $K$ is a constant);
ii) segment $[K t_c,p^{-1}]$; iii) segment $[p^{-1},cn]$ (where $c$ is a constant); iv) segment
$[cn,n (q_{\infty} - \delta)]$. Note that segment i) contains the most crucial, 
initial phase of the process. 

\begin{figure*}[t]
\centering
\includegraphics[width=16cm]{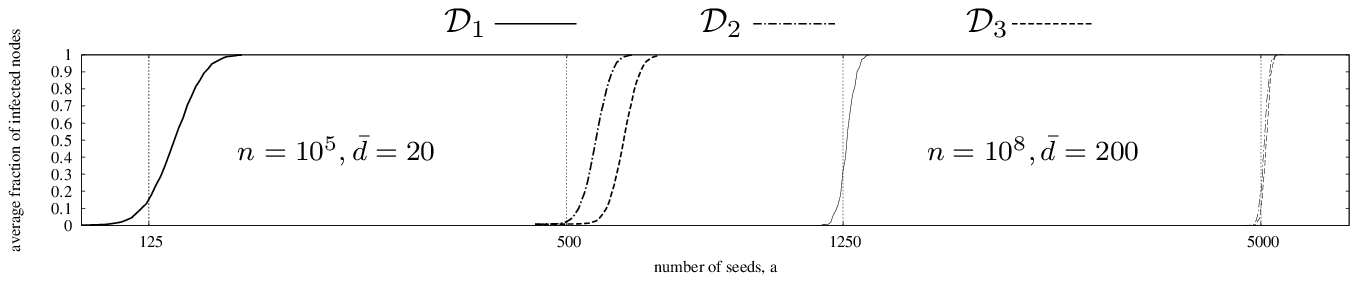}
\vspace{-2mm}
\caption{Phase transitions in $G(n,p)$ for different 
threshold distributions ${\cal D}_1$, ${\cal D}_2$, ${\cal D}_3$, averaging 
the results of $10,000$ simulations. Analytical predictions are shown as vertical
dotted lines.
\label{fig:figura1bis}}
\vspace{-2mm}
\end{figure*}

The following lemma provides a fundamental property related to segment i), which 
provides the key to obtain the result in Theorem \ref{thm:bpercolationsuper}: 
\begin{Lemma} \label{le:add}
Under the assumptions of Theorem \ref{thm:bpercolationsuper}, let
$K>\alpha(1-(\rho^*)^{-1})$ be an arbitrarily fixed constant. 
\begin{equation} \label{senza-modulo}	
\PP\left(\inf_{t\in [a,K t_c]}\{a - t + S(t)\}\le 0\right)=  O\left(\mathrm{e}^{-C_1(\rho^*,\alpha)a+o(a)}\right), \nonumber
\end{equation}
where $C_1(\rho^*,\alpha)$ is given in the statement of Theorem \ref{thm:bpercolationsuper}.
\end{Lemma}

The detailed proof is reported in Appendix \ref{app:lemma}.  
We outline here the three main ingredients to prove Lemma \ref{le:add}:
i) we exploit standard concentration results for the binomial
distribution, providing exponential bounds to $P(|S(t)-\mathbb E[S(t)]|>\epsilon t_c)$ at 
any $t$ in the considered  domain; ii) we employ the union bound to upper bound the probability 
$P(\sup_t |S(t)-\mathbb E[S(t)]|>\epsilon t_c)$ by $\sum_t P(| S(t)-\mathbb E[S(t)]|>\epsilon t_c)$;
iii) we use the property $E[S(xt_c)]=x^{\rho^*}\frac{t_c}{\rho^*}+o(t_c)$.


We emphasize that in this paper we employ different techniques with 
respect to those used in \cite{JLTV}, where authors rely on  
concentration results for $\sup_t |S(t)-\mathbb E[S(t)]|$ 
derived from Martingale theory (Doob's inequality). Instead, we combine 
deviation bounds specifically tailored to the binomial distribution
(see Appendix \ref{app:lemma}) with the union bound, 
obtaining a conceptually simpler approach which also permits us to 
obtain explicit exponential laws for probabilities related to the final 
size of the epidemics (i.e., a stronger result with respect to 
main Theorem 3.1 in \cite{JLTV}, which holds just w.h.p.). 

As immediate consequence of Lemma \ref{le:add} we can say that
the process does not stop before  $K t_c$ with probability 
$1-\zeta(n)$, being $\zeta(n)=O\left(\mathrm{e}^{-C_1(\rho^*,\alpha) a+o(a)}\right)$.

Considering that $E[S(xt_c)] \sim x^{\rho^*}\frac{t_c}{\rho^*}$ quickly (super-linearly) increases 
after $K t_c$ (as long as approximation \equaref{pi-summary} holds), we can expect that
the process is extremely unlikely to stop in segment $[K t_c,p^{-1}]$, if it
survives the first bottleneck segment. The proof of this fact is reported in
Appendix \ref{app:lemma}, where we also handle segment $[p^{-1},cn]$.

Here we focus instead on the last temporal segment, where the value 
of $q_{\infty}$ comes into play determining the final size of the epidemics. Indeed, we are going to 
show that $q_{\infty} n+o(n)$ are infected with probability  $1-\zeta(n)$.
In general, we can assume that $q_\rho = q_\infty - v(\rho) $ with $v(\rho)\to 0$.
Given an arbitrary $\hat{\rho}$ such {\color{black} 
that $ \rho^*\le \hat \rho< cn p$}, 
we make use of concentration inequality (\ref{sotto}) to write:
\begin{align}
\pi(cn)&\geq\sum_{\rho=\hat\rho}^{cn}\binom{cn}{\rho}p^\rho (1-p)^{cn-\rho}q_\rho\nonumber\\
&\ge \sum_{\rho=\hat\rho}^{cn}\binom{cn}{\rho}p^\rho (1-p)^{cn-\rho}
(q_\infty-v(\hat \rho))\nonumber\\
&=\PP(\mathrm{Bin}(cn,p)\geq \hat\rho)(q_\infty-v(\hat \rho)) \nonumber\\
&\geq (q_\infty-v(\hat \rho))(1-\mathrm{e}^{-cn p H(\hat\rho/(cn p))}) \nonumber\\
& \ge q_\infty - (v(\hat \rho)+ \mathrm{e}^{-cn p H(\hat\rho/(cn p))})\nonumber\\
&\ge q_{\infty}-\frac{\epsilon}{2} \label{eq:lim0add-summario}
\end{align}
for any arbitrary $\epsilon>0$ (and $n$ large enough). We have:
\begin{align}
& \hspace{-7mm}  \PP\left( \inf_{t\in[cn,n(q_{\infty}-\epsilon)]} \!\! a-t+S(t) \leq 0\right) \leq
\PP(S(cn)+a \leq
n(q_{\infty}-\epsilon))  \nonumber \\
& \le \PP\left(\mathrm{Bin}(n,1-\pi(cn)) 
\geq n(1-q_{\infty}+\epsilon)\right) \nonumber
\end{align} 
Exploiting (\ref{sopra}), the above probability
goes to 0 faster than $\zeta(n)$ for any $\epsilon>0$, 
proving that at least \mbox{$n(q_\infty - \epsilon)$} nodes are infected.
When $q_{\infty}<1$, we can similarly show that no more
than $n(q_\infty + \epsilon)$ nodes are infected. 
Indeed, considering that $\pi(n(q_\infty +\epsilon)) < q_\infty$, we can apply (\ref{sotto}) 
to show that $\PP(S(n(q_\infty +\epsilon) + a - n(q_\infty +\epsilon) < 0)$ goes to 
1 faster than $\zeta(n)$, for any $\epsilon>0$.



\subsection{Validation} \label{subsec:num1}

To validate our analysis, and understand how well asymptotic results can predict 
what happens in large (but finite) systems, we have run Monte-Carlo simulations of our 
generalized bootstrap percolation model. In each run we change both the identity
of the seeds and the structure of the underlying $G(n,p)$ graph.
We compute the average fraction of nodes that become infected, averaging the results
of 10,000 runs.

We first look at the impact of random thresholds, while keeping
equal weight $W_{ij} = 1$ on all edges. We consider three different distributions
of $R_i$: i) constant threshold equal to 2 (denoted ${\cal D}_1$);
ii) uniform threshold in the set $\{2,3,4,5\}$, (denoted ${\cal D}_2$);
iii) two-valued threshold, with $\PP(R_i = 2) = 1/4$ and $\PP(R_i = 10) = 3/4$  (denoted ${\cal D}_3$); 
Note that all three distributions have $\rho^* = 2$, but their expected values are quite different.
Moreover, $q_{\rho^*} = 1$ for ${\cal D}_1$, whereas $q_{\rho^*} = 1/4$ for both 
${\cal D}_2$ and ${\cal D}_3$.

The asymptotic formula for the critical number of seeds gives in this scenario 
$a_c = n/(2 {\bar{d}}^2 q_{\rho^*})$.
We consider either a \lq small' system, in which $n=10^5$, $\bar{d} = 20$, or a \lq large'
system, in which $n=10^8$, $\bar{d} = 200$. 
Results are shown in Fig. \ref{fig:figura1bis} using a log horizontal scale
on which we have marked the values of $a_c$ derived from the asymptotic formula.
We use the same line style for each threshold distribution,
and different line width to distinguish the small system (thick curves) 
from the large system (thin curves).

We make the following observations: i) the position of the phase transition (i.e., the critical number of seeds) 
is well estimated by the asymptotic formula; ii) despite having quite different shapes, 
distributions ${\cal D}_2$ and ${\cal D}_3$ lead asymptotically to the same critical number of seeds,
as suggested by results obtained in the large system, where the corresponding curves are 
barely distinguishable (at $a_c = 5000$); iii) phase transitions become sharper for higher values of the critical number of seeds,
confirming that the probability law by which the process is supercritical/subcritical
depends strongly on $a_c$ itself (as stated in Theorems \ref{thm:bpercolationsuper} and \ref{thm:bpercolation}).

We next move to a scenario in which the threshold is fixed, $R_i = 2$, and we vary
the weights on the edges. We will consider, for simplicity, a simple case in which
the influence exerted between two nodes can take just two values: +1, with probability $z$,
and -1, with probability $1-z$. Note that the average influence, $\EE[W_{ij}] = 2z-1$, 
can even be negative, if we select $z < 1/2$.
In this scenario, we have $\rho^* = 2$, $q_{\rho^*} = z^2$, hence $a_c = n/(2 (\bar{d} z)^2)$.
We consider either a \lq small' system, in which $n=10^5$, $\bar{d} = 20$, or a \lq large'
system, in which $n=10^7$, $\bar{d} = 200$, which produce the same value of $a_c$, for any $z$.

\tgifeps{8}{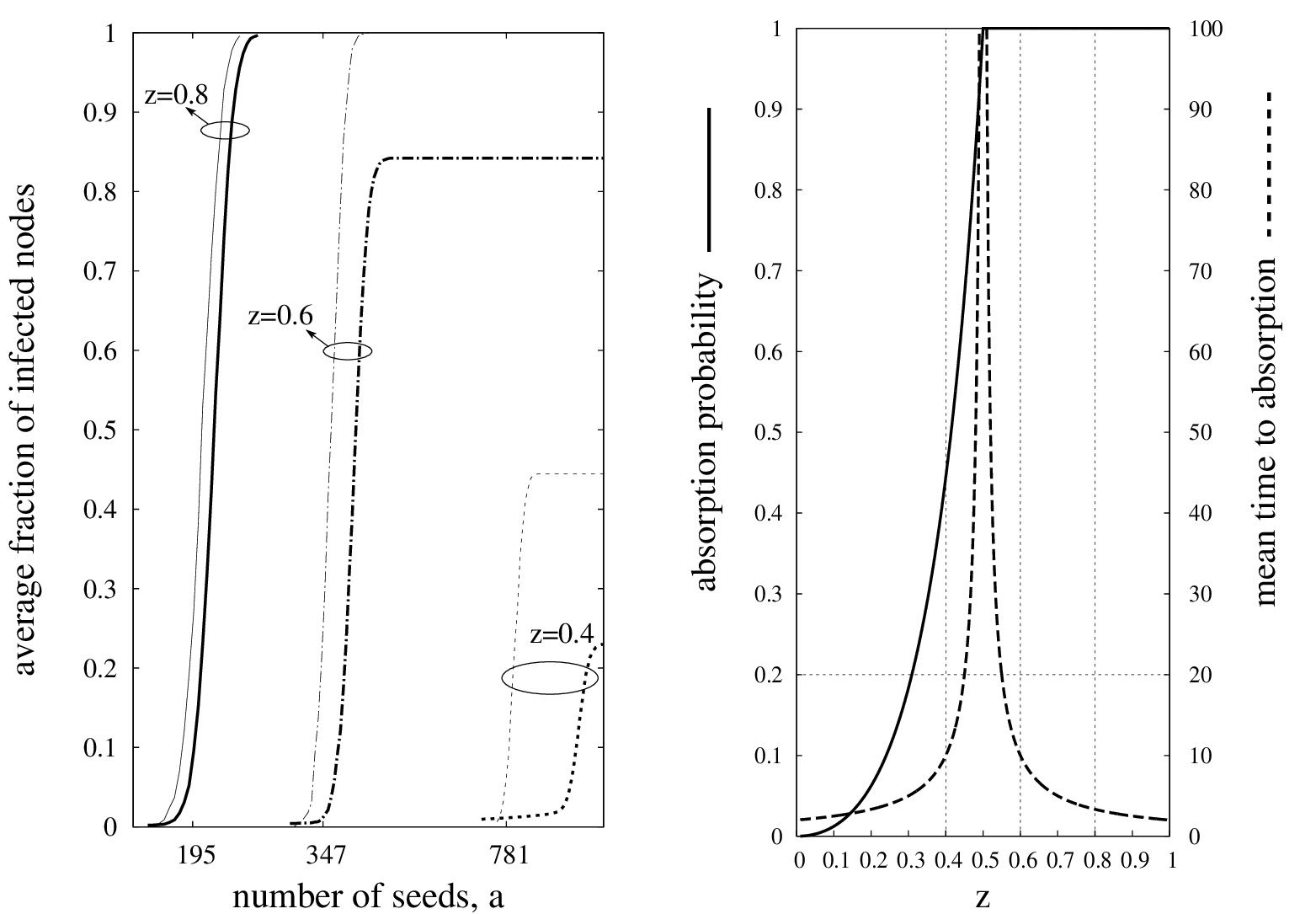}{(left plot) Phase transitions in $G(n,p)$ for fixed threshold $R_i = 2$, 
and random weights $\pm 1$, with $\PP(W_{ij} = 1) = z$. (right plot) results for the 
corresponding simple random walk.}

Results are shown in Fig. \ref{fig:figura2} (left plot), using a log horizontal scale on which we
have marked the values of $a_c$ derived from the asymptotic formula.
We use the same line style for each value of $z$, and different line width 
to distinguish the small system (thick curves) from the large system (thin curves).
We observe that in the small system the average fraction of infected nodes
saturates to a value significantly smaller than one for $z = 0.6$, 
although we expect that, as $n \to \infty$, all nodes should get infected in this case (for which $q_{\infty} = 1$).  
In the large system, the discrepancy between simulation and asymptotic results disappears. 

This phenomenon can be explained by considering that the counter of inactive nodes
behaves as a simple random walk (i.e., with steps $\pm 1$) with an absorbing barrier at $\rho^* = 2$.
Recall \cite{feller} that for this simple random walk the absorption probability is $1$ for $z \geq 1/2$,
while it is equal to $(z/(1-z))^2$ for $z < 1/2$.
Moreover, the mean time to absorption (conditioned to the event that the walk is absorbed)
is $2/|1-2z|$ (see right plot in Fig. \ref{fig:figura2}).
On the other hand, the time horizon of this equivalent random walk is limited by the node degree, since
a node cannot receive a number of contributions to its counter greater than the number of its
neighbors. In the small system the average degree ($\bar{d} = 20$) is too small
to approach the asymptotic prediction, whereas in the large system  
the average degree ($\bar{d} = 200$) is large enough (i.e., much larger than the mean time to 
absorption) to observe convergence of the final size to the asymptotic prediction obtained with 
$q_{\infty}$.
{
Interestingly, a finite fraction of nodes (asymptotically, around 0.44)
gets infected with $z = 0.4$, a case in which the average node-to-node influence
is negative!}


\section{Random graphs with arbitrary degree distribution}\label{sec:degree}
Up to now we have considered the $G(n,p)$ random graph model, and we have followed
the same problem reformulation adopted in \cite{JLTV}, in which 
a single node is used at a time, revealing all its outgoing edges.
This approach is especially suitable to $G(n,p)$,
since marks $M_i(t)$ are i.i.d binomial random variables.
We introduce now an alternative description of the percolation process, 
in which a single edge is used at a time. This approach is more convenient 
to analyze other random graph models, such as $G(n,M)$ (graphs
with pre-established number of edges), $G(n,d)$ (where all nodes
have the same degree), or the configuration model.  

\subsection{Edge-based reformulation for $G(n,M)$} \label{sec:gnm}
We consider the (multi)-graph  $\tilde G(n,M)$ in which,  
starting from a graph with no edges, $M$ edges are sequentially added,
each connecting two nodes selected (independently) uniformly at random. 
Note that by so doing we can generate parallel edges, as well as self loops.
However, following the same approach as in Corollary 3 of \cite{amini1},
it is possible to show that sequences of events that occur  w.h.p.  over 
$\tilde G(n,M)$, occur w.h.p. also over $G(n,M)$, with $G(n,M)$ denoting  
the class of (simple)-graphs having $M$ edges, 
with associated uniform probability law. Therefore our results apply 
to $G(n,M)$ as well.


To analyze bootstrap percolation in $\tilde G(n,M)$, we consider
the following dynamical process: when a node becomes active,
all edges connecting this node to other nodes which are still non active
are denoted as \lq usable', and added to a set $\cB$ of usable edges. 
At a given time step $t$, one usable edge is selected
uniformly at random from  $\cB(t-1)$, adding  one 
mark to {\textcolor{black}{the endpoint that was inactive (when the edge became usable), provided that this endpoint is still inactive}.
The selected edge is then removed from $\cB(t)$. 
Set $\cB(0)$ is initialized with the edges connecting seeds to non-seeds. 
By construction, at most one node can become active at each time instant. 
Hence, denoting with $A(t)$ the number of active nodes at time $t$ (initialized to a), 
we have $A(t) \le a + t$.

Let $\pi(t)$ be the probability that a node, which is not a seed,
has been activated at time $\tau \le t$.   
While it is not easy to write an exact expression of $\pi(t)$, 
we can provide asymptotically tight bounds on $\pi(t)$, as follows:
\begin{align*}
& 1- \sum_{j=0}^{r-1} \binom{t}{j}\Big(\frac {1}{n}\Big)^j\Big(1-\frac{1}{n} \Big)^{t-j} \le
\pi (t) \le  \\
& 1- \sum_{j=0}^{r-1} \binom{t}{j}\Big(\frac {1}{n-a-t}\Big)^j\Big(1-\frac{1}{n-a-t} \Big)^{t-j}
\end{align*}
This because we can reveal the endpoint of an active edge 
only when this edge is used, by choosing uniformly at random one
of the nodes that were non active at the time instant $\tau'$ at which
the considered edge became active. 
Hence, an inactive node $i$ receives a mark at time $\tau$ with probability
$\frac{1}{n-A(\tau')}$ (independently from other previously collected marks). 
Furthermore, by construction, we have 
$\frac{1}{n} \le \frac{1}{n-A(\tau')} \le \frac{1}{n-A(\tau)}\le \frac{1}{n-a-\tau}
\le \frac{1}{n-a-t}$.
At timescale $t=o(n)$, we can approximate $\pi(t)$ as:
\begin{equation}\label{eq:piedges}
 \pi(t)=1-\sum_{j=0}^{r-1} \binom{t}{j}\Big(\frac {1}{n}\Big)^j\Big(1-\frac{1}{n} \Big)^{t-j}+o(1)= 
 \frac{1}{r!}\left(\frac{t}{n}\right)^r+o(1) 
\end{equation}

The dynamics of $\cB(t)$ (whose size is denoted by $B(t)$) obey the following equation:
\[
B(t)=	 B(0) + \Sigma(t)- t
\]
where $\Sigma(t)$ represents the (cumulative) number of edges activated  
at $\tau \le t$. The process stops at time $T = \min\{t: B(t)=0\}$.
Similarly to the $G(n,p)$ case, the number $S(t)$ of nodes that have become active 
by time $t$ is the sum of $n-a$ identically distributed Bernoulli random variables
with average $\pi(t)$. Indeed,  $S(t)=\sum_{i\in \cV \setminus \cA(0)} \ind_{M_i(t)= r}$.

Note that by construction marks are distributed only to inactive nodes, 
therefore a node $i$ stops receiving marks as soon as $M_i(t)=r$.
Differently from $G(n,p)$, however, variables $\ind_{M_i(t)= r}$ are not 
independent, given that at most $t$ marks 
have been distributed by time $t$ (i.e., $\sum_i M_i(t) \le t$). 
Note that we still have $\EE[S(t)]= (n-a) \pi(t)$.   

For what concerns the total number of edges activated by time $t$, $\Sigma(t)$, 
we can express it as the sum of random variables $X_k$ associated
with nodes in $\cA(t)$, representing the numbers of edges activated along with node $k$ 
{\color{black} (i.e. the number of edges connecting node $k$ with inactive nodes)}: 
\[
\Sigma(t)= \sum_{k=1}^{A(t)}X_{k}.
\] 
\textcolor{black}{We can evaluate $X_k$ by dynamically unveiling,
for every inactive edge, whether node $k$ is one of its endpoints (but not both).
It turns out $X_k=\text{Bin}\Big(M-\Sigma(\tau_k-1)-B(0),\\ \frac{2}{n-A(\tau_k-1)}- \frac{2}{(n-A(\tau_k-1))^2}\Big)$
where $\tau_k$ is the time instant at which the $k$-th node was activated. 
Indeed, $M-\Sigma(\tau_k-1)-B(0)$ represents the number of edges still to be activated at time $\tau_k$, 
while $\frac{2}{n-A(\tau_k-1)}-\frac{2}{(n-A(\tau_k-1))^2}$ is
the probability that node $k$ is an endpoint
(but \textcolor{black}{not} both) of any such edges.}
Observe that variables $X_k$ are not independent, as consequence of the fact  
that that sum of all edges in the graph is constrained to be $M$.
However, $X_k$  is conditionally independent from $X_{k'}$, 
with $k'<k$, given $\Sigma(\tau_k-1)$ and $A(\tau_k-1)$.
Moreover, for any $k$ we have:
\begin{align}\label{eq:boundsigma}
\text{Bin}\Big(M-\Sigma(t)-B(0),\frac{2}{n}-\frac{2}{n^2}\Big)  \le_{st}   
X_k \nonumber \\ 
\le_{st}  \text{Bin} \Big(M, \frac{2}{n-a-t}-\frac{2}{(n-a-t)^2} \Big).
\end{align}
In particular, the expectation of $\Sigma(t)$ satisfies:
{\color{black}
\[
\hspace{-3mm} \mathbb{E}\left[\frac{2(n-1)(M-\Sigma(t)-B(0))}{n^2}A(t)\right]  \le \mathbb{E} [\Sigma(t)]\le  \frac{2M\mathbb{E}[A(t)]}{n-a-t}.
\]}
Moreover, under the assumption $a\ll n$, 
since for $t \ll n$, $A(t)\le a + t \ll n $ and {\color{black} $\Sigma(t)+B(0) =o(M)$}, we have: 
\[
\mathbb{E} [\Sigma(t)]= \frac{2M}{n}\mathbb{E}[A(t)](1+o(1))= 2 M\pi(t)(1+o(1))
\]
while $B(0)=\frac{2M}{n}a(1+o(1)))$. Recalling \equaref{piedges},
we have in conclusion:
\[
 \mathbb{E}[ B(t)]=\left(\frac{2M}{n}a+ \frac{2M}{r!}\left(\frac{t}{n}\right)^r -t\right)(1+o(1))
\]

Now, similarly to the case of $G(n,p)$, we can determine the critical number of seeds by: 
i) determining sufficient and necessary conditions under which  { $\mathbb{E}[B(t)]>\delta \frac{M}{n}a$} 
for some arbitrary $\delta>0$ and any $t \ll n$; {\textcolor{black}{so doing we determine the critical number of seeds $a_c$}.
ii) Exploiting the fact that $B(t)$ is sufficiently concentrated around its mean
for $t\le  K \frac{M}{n} a_c $, where $K>1$ is a properly defined constant. 
iii) Showing that for $ K \frac{M}{n} a_c  < t < \ M(1-\epsilon) $,
$B(t)$ can be bounded from below away from 0. 

For what concerns point i) we follow the same lines as for $G(n,p)$, 
defining function $g(t)= 2\frac{M}{n} a +\frac{2M}{r!}\frac{t^r}{n^r}-t$, and 
finding the minimum of $g(t)$, which is achieved at: 
\[
t_c=\left(\frac{(r-1)!n^r}{2M}\right)^\frac{1}{r-1}= 
\frac{2M}{n}\left(\frac{(r-1)!}{\frac{2M}{n} (\frac{2M}{n^2})^{r-1}}\right)^\frac{1}{r-1}
\]
with $t_c=o(n)$ as long as $M\gg n$.
Observe that $\frac{2M}{n}$ is the average node degree
(replacing $np$ in the expression of $t_c$ obtained for $G(n,p)$) while 
$\frac{2M}{n^2}$ can be interpreted as the probability that two specific vertices 
are connected by at least an edge (replacing $p$ for $G(n,p)$).
Evaluating $g(t_c)$ and imposing $g(t_c)=0$, we obtain the critical number of seeds: 
\begin{equation}\label{eq:gnm}
a_c = \left(1 -\frac{1}{r}\right) \left(\frac{(r-1)!}{\frac{2M}{n} (\frac{2M}{n^2})^{r-1}}\right)^\frac{1}{r-1}
\end{equation}
which is exactly the same as what we get in $G(n,p)$ through the substitution 
$\frac{2M}{n}\to np$ and $\frac{2M}{n^2}\to  p$.

For what concerns ii) and iii) we can proceed in analogy with the case of $G(n,p)$,  
exploiting standard concentration results.
In particular, we first focus on time instants $t\le  K\frac{M}{n}a_c$  
for suitable $K>2$. We need to show that $B(t)>0$ w.h.p. provided  
that $\mathbb{E}[B(t)]>\epsilon a_c$ for arbitrary $\epsilon>0$ 
(i.e., $a > (1 + \epsilon) a_c$). To this end observe that from \eqref{eq:boundsigma},
the fact that $\Sigma( K\frac{M}{n}a_c) = o(M)$ and  $A( K\frac{M}{n}a_c) =o(n)$, 
and recalling the above mentioned property of conditional  
mutual independence  of variables $X_{k}$, it descends that w.h.p., for any $t\le  K\frac{M}{n}a_c$: 
$P(\Sigma(t)\le t-B(0))\le P(\sum_{1}^{A(t)} \overline X_k\le  t-B(0))$
with $\overline X_k$ mutually independent and  
$\overline X_k =\text{Bin}\Big(M(1-\epsilon), \frac{2}{n}-  \frac{2}{n^2}\Big)$ for an arbitrarily
small $\epsilon>0$. At last observe that $ P(\sum_{1}^{A(t)} \overline X_k\le  t-B(0))$ can be easily bounded using  
inequalities \eqref{sotto} and \eqref{sopra}.

For what concerns iii) we adopt arguments conceptually similar to the case of $G(n,p)$, 
exploiting the fact that $\mathbb{E}[B(t)]$ quickly (super-linearly) increases after 
$ K\frac{M}{n}a_c$.
 
\subsection{Configuration Model}\label{subsec:CM}
The edge-based problem reformulation described in previous section can be
easily extended to the configuration model $G(n,p(d))$, in which we specify a given
degree sequence (possibly dependent on $n$) with associated empirical distribution 
function $p(d)$. For simplicity, we limit ourselves to describing the computation of the
critical number number of seeds $a_c$. However, the approach can be
made rigorous by following the same lines as for $G(n, M)$.
As before, properties of multi-graphs $\tilde G(n,p(d))$ apply as well to  
simple-graphs $G(n, p(d))$.

Similarly to what we have done for $G(n,M)$, we focus on the evolution of the number of 
activable edges: 
\[
B(t)= B(0) + \Sigma(t) -t 
\]
and compute the critical time $t_c$ by finding the minimum 
of $\EE[B(t)]$.  

The impact of node degree can be taken into account by evaluating the
probability $\pi(t,d)$ that a node with degree $d$ has been activated 
by time $t$. Moreover, we need to consider the amount of edges 
that a node contributes to $\cB$ after being activated.   
There are in total $n \bar{d}$ \lq end-of-edges' in the network, so the probability
that a given end-of-edge is active at time $t$ is $t/(n \bar{d})$.
Hence, we can write: 
\begin{equation}\label{eq:pitd}
\pi(t,d) = 1- \sum_{j = 0}^{r-1}\binom{d}{j} \left(\frac{t}{n \bar{d}}\right)^j 
\left(1-\frac{t}{n \bar{d}}\right)^{d-j} + o(1),
\end{equation}
Since $t/(n \bar{d})$ is small, we can approximate it as
$$ \pi(t,d) = \frac{1}{r!} \left(\frac{d t}{n \bar{d}}\right)^r+ o(1) \qquad (d \geq r) $$
\textcolor{black}{Observe that since, by construction,  a node
gets activated thanks to exactly $r$ active edges,  it contributes $d-r$ 
new edges to $\cB$. Then for $t\ll n$ since $A(t)\le t+a\ll n$  
we can  approximate the average value of $B(t)$ as:}
$$ \mathbb E[B(t)]  \sim B(0) + n \sum_{d \geq r} \frac{1}{r!} \left(\frac{d t}{n \bar{d}}\right)^r (d-r) p(d) -t. $$
Now, if we define 
\begin{equation}\label{eq:dstar}
d^* = \sum_{d \geq r} \left(\frac{d}{\bar{d}}\right)^r \frac{d-r}{\bar{d}} p(d) 
\end{equation}
we obtain $ \mathbb E[B(t)] \sim B(0) + \frac{\bar{d} n t^r}{r! n^r} d^* - t$,
from which we can derive the critical time $t_c$:
$$ t_c  = n \left(\frac{(r-1)!}{\bar{d} \, d^*}\right)^{\frac{1}{r-1}} $$
and the critical number of seeds: 
\begin{equation}\label{eq:acgen}
 a_c  = \left(1 - \frac{1}{r} \right) n \left(\frac{(r-1)!}{\bar{d}^r \, d^*}\right)^{\frac{1}{r-1}} 
\end{equation} 
One can easily check that the above formula is consistent with what we get
in $G(n,p)$ or $G(n,M)$, for which $d^* \sim 1$.
{\color{black} The above formula holds when seeds are selected uniformly at random. 
However, note that our analysis could be immediately extended to the important case 
in which seeds are chosen on the basis of the node degree. Indeed,
what really matters is only the cardinality of the initial set of edges
connecting seeds to non-seeds}. 


\tgifeps{8}{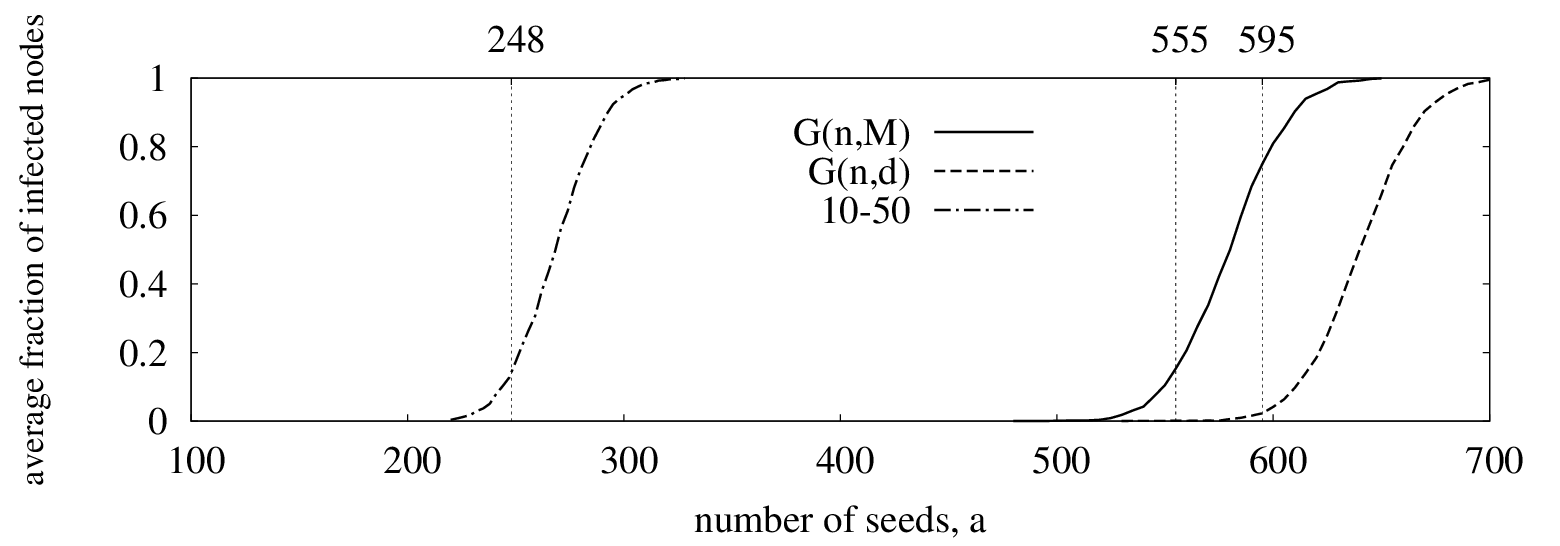}{Phase transitions of basic bootstrap percolation with $r = 2$,
in different random graph models with $n = 10^6$, $\bar{d} = 30$.}

Figure \ref{fig:figura3} reports simulation results for three different random 
graph models having $n = 10^6$ nodes, and average node degree $\bar{d} = 30$.
We consider basic bootstrap percolation with $r = 2$.
We compare the $G(n,M)$ model, the regular $G(n,d)$ (in which nodes have constant degree),
and a configuration model in which half of the nodes have degree 10 while the other half have
degree 50 (curve labelled \lq 10-50'). 
Analytical results obtained by \equaref{gnm} (for $G(n,M)$), and by \equaref{acgen}
(for the other two graph models) are marked on the top margin.
As expected, for fixed average degree, the critical number of seeds 
decreases for increasing variance of the degree distribution.

We experimented also with a real online social network, considering
user-user friendship relations of Orkut, a former
social networking site run by Google. In particular, we 
have used a crawled sub-graph of Orkut with $n= 3,072,441$ nodes \cite{mislove},
hereinafter referred to as Orkut graph,
although it represents only a small percentage (11\%) of the entire social network.
The average node degree of this graph is $\bar{d} = 76.3$, and 
the maximum degree is $d_{\max} = 33,313$.
An interesting question that arises here is the following: 
does a configuration model with the same number of nodes as the Orkut graph, 
and exactly the same degree sequence, produce also a 
similar value of $a_c$? If the answer is affirmative,
it would tell us that the degree distribution alone, and not the entire
network structure, could be used to predict (even analytically) the onset of 
large-scale outbreaks in this kind of systems (as suggested also by \cite{pancolesi}).
We partially answer this question by running simulations on both the original 
Orkut graph and the matched configuration model, as well as by 
analytically evaluating $a_c$ using \equaref{acgen}.

Besides basic bootstrap percolation, we explored also the interesting variation
in which $r$ is a deterministic function of 
the node degree. Indeed, note that \equaref{pitd} can be immediately generalized
to $r = r(d)$, although in this case we do not get a closed-form expression for $a_c$,
and the minimum of $\EE[B(t)]$ has to be computed numerically 
(we omit the details of this computation).
 
\tgifeps{8.5}{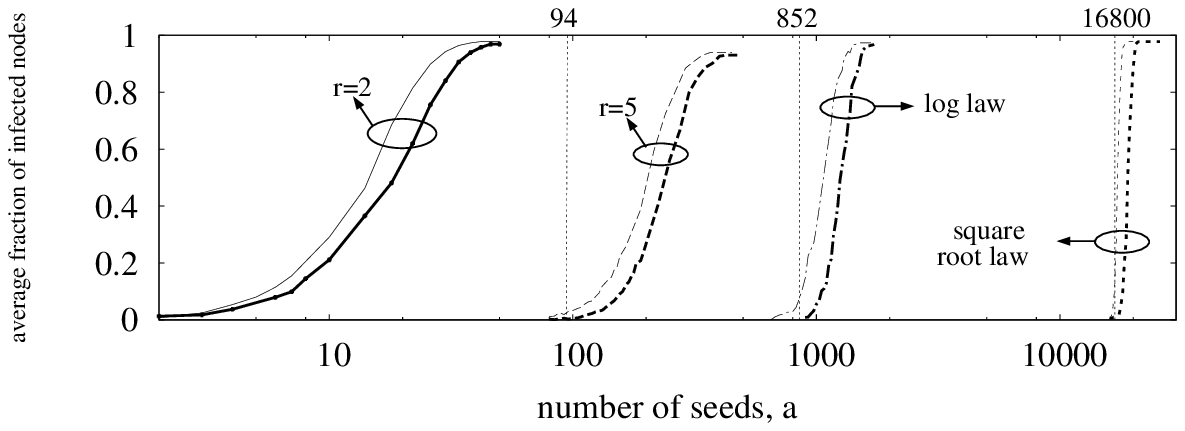}{Phase transitions in the Orkut graph and the matched configuration model, 
for different threshold functions.}
 
Results are shown in Fig. \ref{fig:figura5}, where thick lines
refers to the real Orkut graph, while thin lines
refers to the matched configuration model. We use
different line styles for i) basic bootstrap percolation with $r=2$ or $r=5$;
ii) the logarithmic law $r(d) = \max\{2, \lceil \log_2(d) \rceil\}$;
iii) the square root law $r(d) = \max\{2, \lceil \sqrt{d} \rceil\}$;
Analytical predictions of $a_c$ are shown as vertical dotted lines
(except for $r=2$, for which \equaref{acgen} provides a value of $a_c$
even smaller than $r=2$).
We observe similar phase transitions in the Orkut graph and the associated
configuration model, while the analysis captures quite well the point
after which a major outbreak occurs.

\subsection{Impact of power-law degree distribution}\label{subsec:power}
Large networks observed in a variety of different domains (social, technological, biological networks) 
are characterized by the scale-free property, which implies a power law degree distribution.
Hence, it is interesting to understand the impact of power-law degree distributions
on the critical seed set size.
We will consider here power-law degree distributions of the form
\begin{equation}\label{eq:pd}
p(d) = \frac{C}{d^\beta} \qquad \textrm{for}\,\, d_{\min} \leq d \leq d_{\max}
\end{equation}
where $\beta$ is the power-law 
exponent (typically larger than 2), and $C$ is the normalization factor.
We will further assume that $d_{\max} \rightarrow \infty$, while $d_{\min} = o(d_{\max})$.
Note that by letting $d_{\min}$ scale up with $n$ we can obtain 
an increasing average node degree (graph densification) in the most common
case in which
$\beta > 2$.  

The reason why we introduce a maximum node degree $d_{\max}$ is instead more subtle,
and has to do with the fact that $p(d)$ must be the limiting distribution function
of a sequence of empirical distribution function (for each $n$) associated with 
the configuration model. Clearly, we cannot have in any case a degree larger than $n-1$,
but it turns out that, for the common case of $\beta > 2$, nodes with very large degree     
are so rare 
that is preferable to avoid them at all, setting to zero $p(d)$ after a given $d_{\max} = n^\zeta$, 
with $\zeta < 1$. The maximum value of $\zeta$, for $\beta > 1$, can be obtained
by solving the inequality $\int_{n^\zeta}^{\infty} C x^{-\beta} \diff x > n^{-1}$.
For example, when $\beta > 1$, $d_{\min} = \Theta(1)$, we have $\zeta \leq 1/(\beta-1)$.
In practice, we never see nodes with extremely large degree\footnote{In many real systems
the tail of the degree distribution exhibits an exponential cut-off, and often the degree
cannot exceed a maximum value imposed by physical or technological constraints.}, 
and it is actually customary in many random graph models to assume that the maximum degree is smaller than, say, $n^{1/2}$.   
In our model, we try to be more flexible by allowing a generic $d_{\max} = n^\zeta \ll n$,
satisfying the above constraint (if $\beta > 1$).

In practice, one can starts with a desired distribution $p(d)$
of the form (\ref{eq:pd}), having chosen $d_{\min}$ and $d_{\max}$ (depending on $n$),
and construct a sequence $\{d_i\}_n$ of degrees for the 
configuration model by assigning to node $i$ the degree $d_i = \mbox{inf}\{d: 1-F_n(d) < i/n\}$,
where $F_n(d)$ is the cdf of $p(d)$.
Under our assumptions we have, asymptotically,
$$ C \sim \begin{cases} 
\frac{1-\beta}{d_{\max}^{1-\beta}} &\mbox{if } \beta < 1 \\
\frac{\beta-1}{d_{\min}^{\beta-1}} &\mbox{if } \beta > 1 
\end{cases} $$
The following expression for the generic $k$-th moment of $p(d)$ will come in handy
in the following:
\begin{equation}\label{eq:mompd}
\mathbb E[d^k] \sim \begin{cases} 
d_{\min}^k \frac{\beta-1}{\beta-k-1} &\mbox{if } \beta > k+1 \\
d_{\min}^{\beta-1} d_{\max}^{k+1-\beta} \frac{\beta-1}{k+1-\beta} &\mbox{if } 1 < \beta < k+1 \\
d_{\max}^k \frac{1-\beta}{k+1-\beta} &\mbox{if } \beta < 1
\end{cases} 
\end{equation}
Note that moments of order $k < \beta-1$ depend only on $d_{\min}$ (e.g.,
the average node degree ($k=1$), when $\beta > 2$). Instead,
moments of order $k > \beta-1$ may depend also (or exclusively) on $d_{\max}$.  

Recall that our methodology to compute the critical seed set size requires
that $a_c$ is both $\omega(1)$ and $o(n)$. 
This regime implies that the average node degree $\bar{d}$ cannot be either
too small nor too large. Recall that in the $G(n,p)$ model 
we need that $\bar{d} \gg 1$ and $\bar{d} \ll n^{\frac{r-1}{r}}$.
Under a general degree distribution, it is not stricly necessary that $\bar{d} \gg 1$, 
since (look at formula (\ref{eq:acgen})) we could just have $d^* \rightarrow \infty$, 
resulting into a number of seeds $o(n)$.

To better understand how the critical number of seeds depends
on parameters of the power-law distribution, we evaluate its scaling order with $n$,
assuming for simplicity that $d_{\min} = n^{\gamma}$, with $0 \leq \gamma < \zeta$. 
When $\beta > 1$, we further assume $\zeta \leq \frac{1}{\beta-1}-\gamma$
to avoid rare nodes having very large degree. 
We see from (\ref{eq:dstar}) that $d^*$ depends essentially on the
$(r\!+\!1)$-th moment of $p(d)$, i.e., $d^* = \Theta\left(\frac{\mathbb E[d^{r+1}]}{(\mathbb E[d])^{r+1}}\right)$
(assuming $\mathbb E[d] > r$).
We can thus use the expressions in (\ref{eq:mompd}), and obtain 
that the scaling exponent of $a_c$ is
\footnote{The scaling exponent of a generic function
$f(n)$ is defined as $e(f) := \lim_{n \to \infty} \frac{\log(f(n))}{\log(n)}$.}
\begin{equation}
e(a_c) = \begin{cases} 
1 - \frac{\gamma r}{r-1} &\mbox{if } \beta > r+2, \quad \gamma > 0 \\
1 - \frac{\gamma(\beta-2) + \zeta(r+2-\beta)}{r-1} &\mbox{if } 2 < \beta < r+2 , \quad \gamma \geq 0\\
1 - \frac{\zeta r}{r-1} &\mbox{if } \beta < 2 , \quad \gamma \geq 0
\end{cases} \label{eq:eac}
\end{equation}
\vspace{-3mm}

\tgifeps{8}{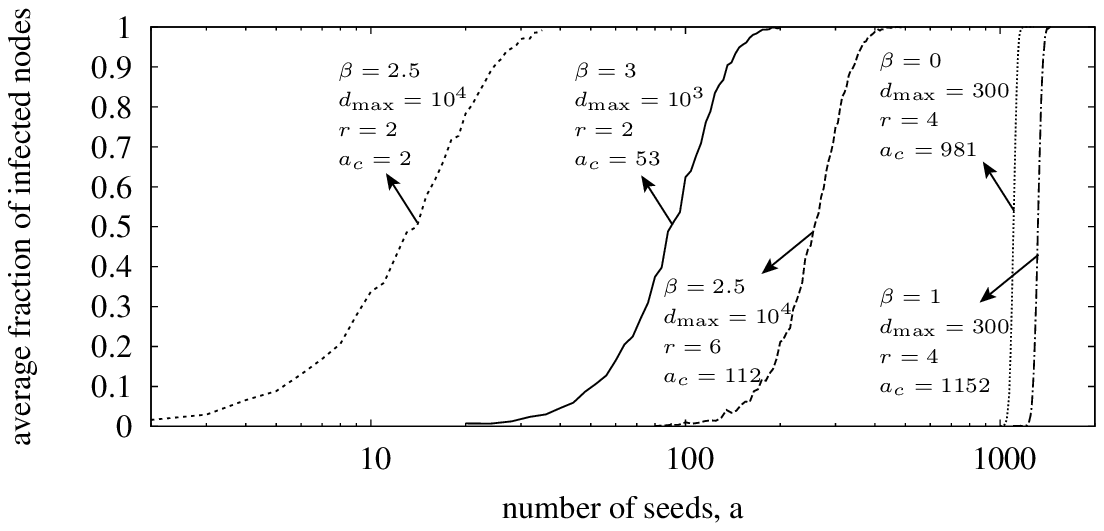}{Phase transitions of basic bootstrap percolation 
in random graphs with $n = 10^6$ nodes and power-law degree.}
\textcolor{black}{We should mention that our results are only partially aligned with those
obtained for Chung-Lu graphs\footnote{\color{black} Interestingly, our scaling exponent
in \equaref{eac}, for $2 < \beta < 3$, $\gamma = 0$, perfectly matches
quantity $a_c^+$ in \cite{amini2} (Theorem 2.3).} with power-law degree distribution in \cite{amini2}, where authors
consider the case $2 < \beta < 3$, $\gamma = 0$.
In particular, in \cite{amini2} they suggest that, when $d_{\max} = \Theta(n^{1/(\beta-1)})$,
$a_c$ is of the order of $n^{\frac{\beta-2}{\beta-1}}$, independently of $r$.}

Figure \ref{fig:figura4} reports simulation results under our power-law
configuration model. The number of nodes is always $n=10^6$, $d_{\min} = 10$, and 
we try different combinations of $\beta$, $d_{\max}$ and $r$.
The values of $a_c$ computed by \equaref{acgen} are also
shown on the plot.
We see that, with $\beta = 2.5$, $d_{\max} = n^{1/(\beta-1)} = 10^4$,
very few seeds are needed with $r=2$, and many more with $r=6$.
We also consider two cases with $\beta < 2$, $d_{\max} = 300$, $r=4$,
to show that, in accordance with \equaref{eac}, when $\beta < 2$,
$a_c$ depends essentially only on the extreme value of the
node degree distribution (i.e., $d_{\max}$), on not on its
shape. Indeed, the phase transitions obtained with $\beta=0$ and $\beta=1$ 
are not that far away, as predicted by our computed values of $a_c$,
despite the fact that the average degree is quite different
in the two cases (i.e., 84 ($\beta=1$) vs 155 ($\beta=0$)).  
 
\section{Community based Graphs: The Block model}\label{sec:block}
Another important feature of many graphs representing real systems
is the presence of a community structure (i.e., a non-negligible
clustering coefficient). This feature is not captured by any of the random
graph models considered so far. In this section, we extend
the analysis of basic bootstrap percolation to the so called block model,
which naturally extends the $G(n,p)$ model to incorporate a 
community structure. We will start from the simple case of just two
communities, and then extend our results to $K < \infty$ communities.

\subsection{The case of two communities}\label{subsec:twocom}
We consider a $G(n_1,n_2,p_1,p_2,q)$ block model comprising
two communities of $n_1$ and $n_2$ nodes, respectively (with $n_1 + n_2 = n$).
The sub-graph induced by nodes belonging to community $i$ (with $i=1,2$)
is an Erd\"os-R\'enyi's graph $G(n_i,p_i)$. Pairs of nodes belonging
to different communities are independently connected with
probability $q$. We assume $q< \min(p_1,p_2)$ 
and $q=\Omega(\max(\frac{1}{n_1}, \frac{1}{n_2}))$.
We denote by $\mathcal{V}_i$ the set of nodes belonging to 
community $i$. 

Bootstrap percolation in $G(n_1,n_2,p_1,p_2,q)$ can be reformulated
in two different ways, which allow us to obtain different (complementary) results.
We explain here our first approach. An alternative reformulation is described
in Appendix \ref{app:second}.

In our first approach we assume that, at each discrete time
step $t$, two active nodes (one in community 1 and one in community 2)
are simultaneously used, whenever they are both available.
If a community runs out of (active) usable  nodes, while the other 
still has some available (active) usable nodes, a single node is used at a time.
We denote by $\mathcal{A}_i(t)$ and $\mathcal{Z}_i(t)$, respectively, 
the set of active nodes and the set of used nodes in community $i$ at time $t$.
Let $A_i(t)=|\cA_i(t)\setminus \cZ_i(t)|$ be the cardinality of the set of 
active usable  nodes in community $i$.
Observe that it is entirely possible that, say, $A_1(t)$ hits zero at some point,
but later on it increases again for effect of marks
received by inactive nodes in $\mathcal{V}_1$ from nodes 
used in $\mathcal{V}_2$. This event makes an exact analysis of the system
particularly difficult.
Note that the process definitely dies at time $T$:
$ T:=\min\{t\in\mathbb N^+:\,\, A_1(t)=0, A_2(t)=0\}$.

We introduce a new quantity $T_i(t)$, representing the number of nodes
that have been used in community $i$ up to time $t$: $T_i(t):=|\cZ_i(t)|$.
From the above discussion, observe that $T_i(t)$ is not necessarily 
equal to $t$, for $t \le T$, in constrast to what happens 
in $G(n,p)$, where $|\cZ_i(t)| = t$, $\forall t \le T$.

The number $S_1(t)$ of initially inactive nodes in $\mathcal{V}_1$
that are active at time $t$ can then be expressed as: 
\[
S_1(t)=\text{Bin}(n-a_1,\widehat{\pi}_i(T_1(t),T_2(t))
\]
where $\widehat{\pi}_{1}(t_1,t_2)=P(\text{Bin}(t_1,p_1)+ \text{Bin}(t_2,q)\ge r)$.
For $p_1t_1\ll 1$   we have:
\[
\widehat{\pi}_{1}(t_1,t_2) \sim \sum_{\rho=0}^{r} \frac{(p_1t_1)^\rho}{\rho!} \frac{(qt_2)^{r-\rho}}{(r-\rho)!}
\] 
(similar expressions can be written for $S_2(t)$ and  $\widehat{\pi}_{2}(t_1,t_2)$ provided that $p_2 t_2\ll 1$).
Note that, whenever $t_1=t_2=t$, previous approximation simplifies to:
$\widehat{\pi}_{1}(t,t) \sim  \sum_{\rho=0}^{r} \frac{p_1^\rho q^{r-\rho}} {\rho!(r-\rho)!}t^r$.
Moreover, if $q\ll p_1$, the latter further simplifies to 
$\widehat{\pi}_{1}(t,t)\sim \frac{ (p_1)^{r}} {r!}t^r$, as in $G(n,p)$.

To characterize the system behavior, we stochastically 
upper and lower bound $S_i(t)$, for $i=1,2$, by two virtual processes
$\underline{S}_i(t)$ and $\overline{S}_i(t)$ obtained
in the following shadow systems:
a shadow reduced system, in which inter-community
edges are removed, and thus each community evolves as in
isolation. 
Note that process $\underline{S}_i(t)$ associated to the reduced system
is equivalent to what we would get in a $G(n_i,p_i)$ model;
a shadow augmented system (viewed by community $i$), in which we assume that 
a new node can {\em always} be used in the other community (if the other community has no usable nodes, 
an arbitrary inactive node in the other community is selected and used).

It immediately descends from their definitions  that:
\begin{eqnarray*}
\underline{S}_1(t)  &=& \text{Bin}(n_1-a_1, \widehat{\pi}_{1}(\underline{T}_1(t),0)) \\
\overline{S}_1(t) &=& \text{Bin}(n_1-a_2, \widehat{\pi}_{1}(\overline{T}_1(t), t))
\end{eqnarray*}
Furthermore, by induction over time, it can be immediately shown that: 
$\underline{S}_1(t)\le_{st} {S}_1(t)  \le_{st} \overline{S}_1(t)$, 
where  $\le_{st}$  indicates the less or equal  
operator under usual stochastic ordering 
(also known as \lq first order stochastic dominance').
Now, under the joint conditions $\mathcal{A}_1(\tau)\neq \mathcal{\cZ}_1(\tau)$ 
and $\mathcal{A}_2(\tau)\neq \mathcal{\cZ}_2(\tau)$ for any $\tau<t$,
by construction $S_1(\tau)=\overline{S}_1(\tau)$ and  
$S_2(\tau)=\overline{S}_2(\tau)$ for any $\tau < t$,
which allows us to conclude that also at time $t$: 
$S_1(t) = \overline{S}_1(t)$ and  $S_2(t) = \overline{S}_2(t)$.
Then, by induction over time:
\begin{align*}
& \left\{\cA_1(\tau)\neq \cZ_1(\tau), \cA_2(\tau)\neq \cZ_2(\tau), \forall \tau\le t \right\}\\ 
=&\left\{\inf_{\tau< t} \min(S_1(\tau)-\tau+a_1,S_2(\tau)-\tau+a_2)>0 \right\}\\
=&\left\{\inf_{\tau< t} \min(\overline{S}_1(\tau)-\tau+a_1,\overline{S}_2(\tau)-\tau+a_2)>0 \right\}.
\end{align*}

In particular, whenever $\overline{S}_1(t)$ and  $\overline{S}_2(t)$ both satisfy 
supercritical conditions, then necessarily  $\cA_i(t)\neq \cZ_i(t)$  
and $S_i(t)=\overline{S}_i(t)$, for any $t<\max(n_1,n_2)-o(\max(n_1,n_2))$.
More formally, exploiting the results in \ref{subsec:main}, we get:
{\color{black}
\begin{Theorem}\label{thm:bpercolationsuper-comunity}
Suppose $1/(n_ip_i)\to 0$ , $p_i=o(n_i^{-1/r})$,  
$a_i/a^{(i)}_c\to\alpha_i$, for some $\alpha_i>1$ with $i=\{1,2\}$,
and: 
\[
 a_c^{(i)}= \left(1-\frac{1}{r}\right)\left(\frac{(r-1)!}{n_i(\widehat{p}_i)^{r}}\right)^{(r-1)^{-1}}
\]
where $\widehat{p}_i= \sqrt[r]{(r!) \cdot \sum_{\rho=0}^{r} \frac{p_i^\rho q^{r-\rho}} {\rho!(r-\rho)!}}$.
Then, having defined $T=\inf_{t}\{\min_i(a_i - t + S_i(t))<0\}$
we have: 
\begin{align*}
\hspace{-3mm} \forall \varepsilon>0,  
P\!\left(\!1\!-\frac{T}{\max_i(n_i)}>\varepsilon\!\right)\!\!=
O\left(\mathrm{e}^{-C_1(r,\min_i(\alpha_i))a+o(a)}\right)\vspace{-2mm}
\end{align*}
where $C_1(r,\alpha)$ is the same function as in Theorem \ref{thm:bpercolationsuper}. 
\end{Theorem}
}

Theorem \ref{thm:bpercolationsuper-comunity} provides sufficient (but not necessary)
conditions for almost complete percolation of $G(n_1,n_2,p_1,p_2,q)$.  

To complement previous result, suppose that $\underline{S}_1(t)$ satisfies supercritical
conditions (while to avoid trivialities we assume $\overline{S_2}(t)$ to be sub-critical).
In this case complete or almost complete percolation occurs in $\mathcal{V}_1$ 
as immediate consequence of Theorem \ref{thm:bpercolationsuper} applied to community $1$
in isolation, and the fact that $\underline{S}_1(t)\le S_1(t)$.  
Then, under the assumption that $q \ge \Omega(\max(\frac{1}{n_1},\frac{1}{n_2}))$,  
we obtain complete or almost complete percolation also in community 2, since
any node in $\mathcal{V}_2$ would have finite probability of having at 
least $r$ neighbors in $\mathcal{V}_1$\footnote{More in general, for  \textcolor{black}{$q\ll \frac{1}{n_2}$}, we could study percolation in community 2 in isolation after: i) adding to $a_2$ the nodes 
in $\mathcal{V}_2$ that have at least $r$ neighbors in $\mathcal{V}_1$; ii) reduce
the threshold $r$ for all inactive nodes in $\cV_2$ to a stochastic 
threshold $R\le r$ accounting for marks received from $\mathcal{V}_1$.}.

It remains to analyze the case in which $\overline{S}_1(t)$ is super-critical (but  $\underline{S}_1(t)$ is sub-critical), and $\overline{S}_2(t)$ is not 
super-critical (or viceversa). This case, which can occur only when $q$ and $p_1$
are of the same order, is more involved and we leave it to future study.

\vspace{-2mm} 
\subsection{Extension to the general block model}\label{sect:blockmodel-ext}
Results obtained for the case of two communities can be rather easily extended to a more general block model with $K<\infty$ heterogeneous communities, specified by a symmetric matrix $P$,
whose element $p_{jk}$ represents the probability that 
a node belonging to community $j$ is connected to a node
belonging to community $k$ (with i.i.d. probabilities for 
all such pairs). Note that diagonal elements of $P$ provide inter-community
edge probabilities. In Appendix \ref{app:blockmodel-ext},
besides our system assumptions, we present two different upper bounds 
on the critical number of seeds. Interestingly, both bounds 
depends critically on extreme values of the model parameters,
and in some cases a single community can determine the phase transition 
of the entire system.

\vspace{-2mm}
\section{Conclusions}
We proposed a unique framework to study sub-linear
phase transitions of threshold-based \lq activation processes'
in random graphs, adding inhomogeneities in the system 
along different (orthogonal) directions. We found that in several cases the 
critical seed set size depends critically just on extreme values of 
distributions, providing novel insights into the dynamics
of epidemic processes in complex systems.

\vspace{-3mm}
\begin{small}
\bibliographystyle{abbrv}

\end{small}

\appendix

\section{Theorem 4.1 
}\label{app:lemma}


First we introduce the following concentration inequalities
for the binomial distribution $\text{Bin}(n,p)$, taken from \cite{P}:
Let $H(b)=1-b+b\log b$, for $b>0$. Let $\mu=n p$.

If $k\le \mu$ then: \vspace{-0mm}
\begin{equation} \label{sotto}
\PP(\text{Bin}(n,p))\le k)\le \exp \left(-\mu H\left(\frac{k}{\mu}\right)\right)
\end{equation} \vspace{-1mm}

If $k> \mu$ then:  \vspace{-1mm}
\begin{equation} \label{sopra}
\PP(\text{Bin}(n,p))\ge k)\le \exp \left(-\mu H\left(\frac{k}{\mu}\right)\right)
\end{equation} \vspace{-0mm}

\vspace{2mm} 
{\it Proof of Lemma \ref{le:add}.}

For  any $\delta>0$,
We have $a=\alpha\,a_c + o(1) >(\alpha-\delta)a_c$, and so, by the definition of $a_c$, for any $t=a,\ldots,\lfloor K t_c\rfloor$, 
\begin{align}
\left\{S(t)- t + a  \le 0 \right\}  \subseteq &
\left\{ S(t) -t +(\alpha-\delta)(1-(\rho^*)^{-1})t_c  \le 0 \right\} \nonumber\\
&=\left\{ S(t)  \le t -(\alpha-\delta)(1-(\rho^*)^{-1})t_c  \right\} \nonumber
\end{align}
\vspace{-1mm}
Hence: 
\begin{align} \vspace{-1mm}
&\left\{ \inf_{t [a,\lfloor Kt_c \rfloor]}S(t)-t+a\le 0\right\} = \bigcup_{t\in[a,\lfloor Kt_c \rfloor] }\{S(t)-t+a\le 0\} \nonumber\\
& \subseteq \bigcup_{t\in[a,\lfloor Kt_c \rfloor] }\left\{ S(t)  \le t -(\alpha-\delta)(1-(\rho^*)^{-1})t_c  
\right\} \nonumber
\end{align}
Moreover, defining $x=t/t_c$, it can be be proved that:
\vspace{-1mm}
\begin{equation}\label{eq:mediaSxtc}
E[S(xt_c)]=x^{\rho^*}\frac{t_c}{\rho^*}+o(t_c).
\end{equation}
\vspace{-1mm}
Using \equaref{mediaSxtc} and \eqref{sotto}, we have for any $\delta>0$, 
\begin{align*}
 &P( S(t)  \le t -(\alpha-\delta)(1-(\rho^*)^{-1})t_c)< \\
 & \mathrm{e}^{ -(1-\delta) x^{\rho^*}\frac{t_c}{\rho^*}
 H\left(\frac{x t_c -(\alpha-\delta)(1-(\rho^*)^{-1})t_c }{(1-\delta) x^{\rho^*}\frac{t_c}{\rho^*} } \right) }
\end{align*}
Thus, by sub-additivity of probability:
\begin{align}
&P\left( \inf_{t [a,\lfloor Kt_c \rfloor]}S(t)-t+a\le 0\right) \le \nonumber \\
& \sum_{t\in [a,\lfloor Kt_c\rfloor]} \mathrm{e}^{ -(1-\delta) x^{\rho^*}\frac{t_c}{\rho^*}
 H\left(\frac{x t_c -(\alpha-\delta)(1-(\rho^*)^{-1})t_c }{(1-\delta) x^{\rho^*}\frac{t_c}{\rho^*} } \right) } \nonumber \\  
 & \le  \frac{K}{\alpha (1-(\rho^*)^{-1})(1+\delta)} \nonumber \\ 
 &\hspace{-3 mm} \mathrm{e}^{ -\inf_{x\in[a/t_c,K]}\left[\frac{(1-\delta) x^{\rho^*}a}{\alpha (1-(\rho^*)^{-1})\rho^*(1+\delta)}
 H\left(\frac{x  -(\alpha-\delta)(1-(\rho^*)^{-1}) }{(1-\delta) \frac{x^{\rho^*}}{\rho^*} } \right)\right]} \label{M3-final}
 \end{align}
the assertion descends immediately taking the inf of \eqref{M3-final} with respect to $\delta>0$ and letting $K\to \infty$.

\vspace{4mm}
{\it Segment $[Kt_c,p^{-1}]$.}
We basically follow \cite{JLTV}, choosing $K=8$ and defining a sequence of time instants 
$t_j=8\cdot{2^j}t_c$  for $j=\{0,1,2\cdots J\}$ with $J=\min\{j: pt_j\ge 1\}$. 
We first show that   $ \mathbb E[S(t_j)]>(1+\delta)t_{j+1}$ for every $j$ and a properly specified  $\delta>0$.
Then, applying again union bound and concentration inequality \eqref{sotto}  we can prove 
that $P(S(t_j)\le t_{j+1}, \text{ for some } j)$ goes to zero faster than 
 $\zeta(n)$. This  implies  $P(S(t)-t \le 0, \text{ for some } t \in [ 8t_c, \lceil p^{-1}\rceil])$ goes to 0 
faster than $\zeta(n)$ under super-critical conditions.  
Indeed, given the monotonicity of $S(t)$,  we have  $\{a-t + S(t)<0 \text{ for some } t\in [t_j, t_{j+1}]  \}
\subseteq \{S(t_j)<t_{j+1}\}$. In conclusion, under super-critical conditions the 
process never stops before $p^{-1}$ with probability $1-\zeta(n)$. 

\vspace{4mm}
{\it Segment $[p^{-1},c\,n]$.}
Beyond time $p^{-1}$ we can no longer use \equaref{pi-summary}. However,
we can easily handle segment $[p^{-1},cn]$ and already conclude that, in all cases, the process
reaches at least a constant fraction of the nodes (if it survives the bottleneck).
For this, we exploit the fact that $\pi(\lceil p^{-1} \rceil)\ge q_{\rho^*}P(\text{Bin}(p^{-1},p)>R_i) >2c$ 
for some constant $c>0$. Using again union bound and concentration inequalities,
we then show that the process never stops before $cn$ with probability $1-\zeta(n)$.


\section{Alternative approach for two communities}\label{app:second}
We introduce a different approach for the $G(n1,n2,p_1,p_2,q)$ block model, which 
allows us to understand how seeds should be optimally
partitioned between the two communities in order to minimize
their number and achieve almost complete percolation in the whole system. 
This time, we assume that at each time step $t$ a single active node, 
selected uniformly at random among all \textcolor{black}{usable} active nodes in the system, is used.

To simplify the exposition, we will focus on a perfectly
symmetric scenario in which $p_1=p_2$ and $n_1=n_2=n/2$. However,
the same approach can be easily extended to the general  
case $G(n_1,n_2,p_1,p_2,q)$.

Differently from our first reformulation, now we have, for any $t < T$:
\[
T_1(t)+ T_2(t)=t
\]
where $T_i(t)$ denotes the (random) number of nodes used
in community $i$ up to time $t$.

Now, if we consider any two different sequences in which active nodes
are selected, such that $T_1(t)$ in one sequence is
larger than $T_1'(t)$ in the other sequence,
we easily see that,
\begin{eqnarray}
&S_1(T_1(t),T_2(t))\ge_{st}S_1(T_1'(t),T_2'(t)) \label{eq:s1} \\ 
&S_2(T_1(t),T_2(t))\le_{st}S_2(T_1'(t),T_2'(t)) \label{eq:s2}
\end{eqnarray}

Furthermore, for $1 \ll t \ll p^{-1}$ we have:
\begin{align} \label{eq:Stotal}
&\mathbb{E}[S_1(T_1(t),T_2(t))+S_2(T_1(t),T_2(t))]\sim \nonumber\\
& \hspace{-0mm} \sum_{\rho=0}^r\!{\frac{( pT_1(t))^\rho(qT_2(t))^{r-\rho}}{\rho!(r-\rho)!} }+\!
\sum_{\rho = 0}^r\!{\frac{( qT_1(t))^\rho(pT_2(t))^{r-\rho}}{\rho!(r-\rho)!} } = \nonumber \\
& f\left(\!\frac{T_1(t)}{t}\!\right)
\end{align}
being $f(x)=\sum_{\rho}[{\frac{( ptx)^\rho(qt(1-x))^{r-\rho}}{\rho!(r-\rho)!} } +{\frac{( qtx)^\rho(pt(1-x))^{r-\rho}}{\rho!(r-\rho)!} }]$ \\
a continuous function over $[0,1]$, indefinitely derivable in $(0,1)$  
and satisfying the following properties: i) $f(x)=f(1-x)$; ii) $f(x)$ decreases  
for $x \in [0,1/2)$ (and increases for $x\in (1/2,1]$).
Previous observations lead to:
\begin{Theorem} \label{seeedsincommunities}
To minimize the number of seeds that are needed to achieve complete or 
almost complete percolation in the symmetric graph $G(n/2,n/2,p,p,q)$, with $q<p$ 
and $q =\Omega(\frac{1}{n})$, all seeds have to be placed within the same community.
\end{Theorem}
The proof is reported in the companion technical report ~\cite{techrep}.
At high level, the result descends from the fact that, for any given \textcolor{black}{total} number of seeds,
extremal trajectories of $T_1(t)$ and $T_2(t)$ are obtained when all seeds 
are placed in the same community, as \textcolor{black}{a} consequence of \equaref{s1}, \equaref{s2}, and
properties of ~\equaref{Stotal}.

 
Theorem \ref{seeedsincommunities} can be easily generalized to the asymmetric case
(see \cite{techrep}):
\begin{Theorem} \label{seeedsincommunitiesasymp}
To minimize the number of seeds that are needed to achieve complete or 
almost complete percolation in $G(n_1,n_2,p_1,p_2,q)$ with 
$q =\Omega(\frac{1}{n})$, all seeds have to be placed 
in the community having the maximum value of $n_i(p_i)^r$. 
\end{Theorem}

 

\section{General block model}\label{app:blockmodel-ext}
Let $n_k(n)$ be the number of nodes in community $k$ ($k = 1,\dots,K$), with $n=\sum_k n_k$. 
We will assume that $n_k(n)\gg 1$, for any $k$.
We focus on a community structure in which $p_{ik}<\min(p_{ii},p_{kk})$ for any $(i,k)$. 
Moreover, whenever $p_{ik}\neq 0$, we will assume
that $p_{ik}=\Omega(\frac{1}{n_i},\frac{1}{n_k}$). 
At last, but without loss of generality, we assume 
the graph to be connected at the community level; 
i.e. we assume $P$ to be of maximal rank (equal to $K-1$). 
Indeed, if this in not true we can always partition the community-level 
graph into connected components and apply our results 
to each connected component.  
 
We first generalize the result in Theorem \ref{thm:bpercolationsuper-comunity}:
\begin{Theorem}\label{thm:bpercolationsuper-comunity-extended}
Consider a block model with $K<\infty $ communities as defined before;
suppose, for any $k$, that $1/(n\,p_{k,k})\to 0$ , $p_{k,k}=o(n^{-1/r})$,
$a_k/a^{(k)}_c\to\alpha_k > 1$, with: 
\[
\bar a_c^{(k)}= \left(1-\frac{1}{r}\right)\left(\frac{(r-1)!}{n_{k}(\widehat{p}_{k,k})^{r}}\right)^{(r-1)^{-1}}
\]
where: \vspace{-5mm} 
\[
\widehat{p}_{k,k}= \left( (r!)\cdot \sum_{\stackrel{\rho_1 \cdots \rho_k \cdots \rho_K} {\text{with } \sum  \rho_j=r}}
\frac{ p_{k,k}^{\rho_k} \prod_{j\neq k} p_{jk}^{\rho_j} }
{\rho_k! \prod_{j\neq k} \rho_j ! }\right)^{\frac{1}{r}}
\]
Let $T=\inf_{t}\{\min_k(a_k - t + S_k(t)) < 0\}$. We have:
\[
\hspace{-3mm} \forall \varepsilon>0,  
P\!\left(\!1-\!\frac{T}{\max_k(n_k)}>\varepsilon\right)\!=O\!\left(\mathrm{e}^{-C_1(r,\min_k(r,\alpha_k))a+o(a)}\right)
\]
where $C_1(r,\alpha)$ is the same function as in Theorem \ref{thm:bpercolationsuper}.
\end{Theorem}

Theorem \ref{thm:bpercolationsuper-comunity-extended} can be used to derive
a simple upper bound to the minimum number of seeds that
can produce super-critical conditions in all communities, 
in the case in which seeds are selected uniformly 
at random among all nodes. Indeed, Theorem \ref{thm:bpercolationsuper-comunity-extended} 
coupled with standard concentration arguments lead to the result that a global number of seeds:
\begin{equation} \label{upper1}
a=(1+\epsilon)n \max_k  \frac{\bar a_c^{(k)}}{n_k}
\end{equation}
for any $\epsilon>0$, is enough to guarantee 
(almost) complete percolation of the entire graph.

More in general, given an arbitrary allocation of seeds among communities,
Theorem \ref{thm:bpercolationsuper-comunity-extended} can be used
to check whether the considered seed allocation is able to trigger system-wide
percolation. On this regard, note that Theorem \ref{thm:bpercolationsuper-comunity-extended}
can be applied to any community-level connected sub-graph of the entire system: if
at least one sub-graph satisfies the conditions of Theorem \ref{thm:bpercolationsuper-comunity-extended}, we get (almost) complete percolation of the entire system, as
consequence of the assumptions that: i) the graph is connected at community level; 
ii) for non null off diagonal elements of $P$, $p_{ik}=\Omega(\frac{1}{n_i},\frac{1}{n_k})$.
In particular, note that if we get (almost) complete percolation in just one
community, the infection propagates to the entire system.

We can also ask ourselves which is the optimal seed allocation in the system,
in the case in which we know the community membership of the nodes.
A straightforward extension of Theorem \ref{seeedsincommunitiesasymp}
provides the answer to this question:
\begin{Theorem} \label{seeedsincommunitiesasympext}
Consider a general block model graph with $K<\infty $ communities as defined before;
assume that  $1/(np_{k,k})\to 0$ , $p_{k,k}=o(n^{-1/r})$ for any $k$.
In order to minimize the number of seeds that produce (almost) complete 
percolation of the entire graph, all seeds must be placed
within a single community that maximizes quantity $n_k(p_k)^r$.  
\end{Theorem}

As consequence, it turns out that a number of seeds:
\begin{align} \label{eq:a2}
a = (1+\epsilon)\left(1-\frac{1}{r}\right)\min_k\left(\frac{(r-1)!}{n_{k}(p_{k,k})^{r}}\right)^{(r-1)^{-1}}
\end{align}
for any $\epsilon>0$, is enough to guarantee an almost complete percolation of 
the graph. Indeed, by placing these $a$ nodes all within 
a single community that maximizes quantity $n_k(p_k)^r$, let this community 
be $k_0$, we have that process $\underline S_{k_0}(t)$, and thus
process $S_{k_0}(t)$ is super-critical, which is enough to
trigger system-wide percolation.

At last, we can exploit Theorem \ref{seeedsincommunitiesasympext}
also to get a different upper bound to the critical number of seeds
in the case in which seeds are selected uniformly at random among all nodes.
Indeed, it turns out that $a \frac{n}{n_{k_0}}$ seeds, where $a$ is the same
as in \equaref{a2}, are enough, since w.h.p. at least $a (1-\varepsilon/2)$ 
seeds will fall within community $k_0$, producing super-critical conditions
in community $k_0$ (and then in the entire system).
In the case of very heterogeneous communities, this last bound
might be tighter than \eqref{upper1}.

\end{document}